\documentclass[preprint]{aastex62}
\usepackage{graphicx}
\usepackage{color}

\received{17 April 2019}
\revised{26 May 2019}
\accepted{16 June 2019}

\submitjournal{ApJ}

\shorttitle{Impact of Stellar Superflares on Planetary Habitability}
\shortauthors{Yamashiki et al.}

\begin{document}

\title{Impact of Stellar Superflares on Planetary Habitability}

\correspondingauthor{Yosuke A. Yamashiki}
\email{yamashiki.yosuke.3u@kyoto-u.ac.jp}

\author{Yosuke A. Yamashiki}
\affil{Graduate School of Advanced Integrated Studies in Human  Survivability, Kyoto University, Sakyo, Kyoto, Japan}
\affil{Unit of the Synergetic Studies for Space, Kyoto University, Sakyo, Kyoto, Japan}

\author[0000-0003-0332-0811]{Hiroyuki Maehara}
\affil{Okayama Branch Office, Subaru Telescope, National Astronomical Observatory of Japan, NINS, Kamogata, Asakuchi, Okayama, Japan}
\affil{Okayama Observatory, Kyoto University, Kamogata, Asakuchi, Okayama, Japan}

\author[0000-0003-0332-0811]{Vladimir Airapetian}
\affil{NASA/GSFC/SEEC, Greenbelt, MD, USA}
\affil{American University, DC, USA}

\author[0000-0002-0412-0849]{Yuta Notsu}
\affil{Laboratory for Atmospheric and Space Physics, University of Colorado Boulder, Boulder, CO, USA}
\affil{National Solar Observatory, Boulder, CO, USA}
\affil{Department of Astronomy, Kyoto University, Sakyo, Kyoto, Japan}

\author{Tatsuhiko Sato}
\affil{Nuclear Science and Engineering Center Center, Japan Atomic Energy Agency (JAEA), Tokai, Ibaraki, Japan}

\author[0000-0003-2493-912X]{Shota Notsu}
\affil{Leiden Observatory, Leiden University,  Leiden, The Netherlands}
\affil{Department of Astronomy, Kyoto University, Sakyo, Kyoto, Japan}

\author{Ryusuke Kuroki}
\affil{Graduate School of Advanced Integrated Studies in Human Survivability, Kyoto University, Sakyo, Kyoto, Japan}

\author{Keiya Murashima}
\affil{Faculty of Science, Kyoto University, Sakyo, Kyoto, Japan}

\author{Hiroaki Sato}
\affil{Faculty of Engineering, Kyoto University, Sakyo, Kyoto, Japan}

\author[0000-0002-1297-9485]{Kosuke Namekata}
\affil{Department of Astronomy, Kyoto University, Sakyo, Kyoto, Japan}

\author[0000-0003-1242-7290]{Takanori Sasaki}
\affil{Department of Astronomy, Kyoto University, Sakyo, Kyoto, Japan}
\affil{Unit of the Synergetic Studies for Space, Kyoto University, Sakyo, Kyoto, Japan}

\author{Thomas B. Scott}
\affil{Interface Analysis Centre, University of Bristol, Bristol, UK}

\author{Hina Bando}
\affil{Faculty of Science, Kyoto University, Sakyo, Kyoto, Japan}

\author{Subaru Nashimoto}
\affil{Faculty of Science, Kyoto University, Sakyo, Kyoto, Japan}

\author{Fuka Takagi}
\affil{Faculty of Agriculture, Kyoto University, Sakyo, Kyoto, Japan}

\author{Cassandra Ling}
\affil{Graduate School of Advanced Integrated Studies in Human Survivability, Kyoto University, Sakyo, Kyoto, Japan}

\author{Daisaku Nogami}
\affil{Department of Astronomy, Kyoto University, Sakyo, Kyoto, Japan}
\affil{Unit of the Synergetic Studies for Space, Kyoto University, Sakyo, Kyoto, Japan}

\author{Kazunari Shibata}
\affil{Astronomical Observatory, Kyoto University, Sakyo, Kyoto, Japan}
\affil{Unit of the Synergetic Studies for Space, Kyoto University, Sakyo, Kyoto, Japan}

\begin{abstract}
High-energy radiation caused by exoplanetary space weather events from planet-hosting stars
can play a crucial role in conditions promoting or destroying habitability in addition to the conventional factors. 
In this paper, we present the first quantitative impact evaluation system of stellar flares on the habitability factors with an emphasis on the impact of Stellar Proton Events. 
We derive the maximum flare energy from stellar starspot sizes and examine the impacts of flare associated ionizing radiation on CO$_2$, H$_2$, N$_2$+O$_2$ --rich atmospheres
of a number of well-characterized terrestrial type exoplanets.
Our simulations based on the Particle and Heavy Ion Transport code System [PHITS] suggest that the estimated ground level dose for each planet in the case of terrestrial-level atmospheric pressure (1 bar) for each exoplanet does not exceed the critical dose for complex (multi-cellular) life to persist, even for the planetary surface of Proxima Centauri b, Ross-128 b and TRAPPIST-1 e. 
However, when we take into account the effects of the possible maximum flares from those host stars, the estimated dose reaches fatal levels at the terrestrial lowest atmospheric depth on TRAPPIST-1 e and Ross-128 b. 
Large fluxes of coronal XUV radiation from active stars induces high atmospheric escape rates from close-in exoplanets suggesting that the atmospheric depth can be substantially smaller than that on the Earth. In a scenario with the atmospheric thickness of 1/10 of Earth's, the radiation dose from close-in planets including Proxima Centauri b and TRAPPIST-1 e reach near fatal dose levels  with annual frequency of flare occurrence from their hoststars. 
\end{abstract}

\section{Introduction}
The definition of habitable zones for extrasolar planetary systems is traditionally based on the conditions promoting the presence of standing bodies of liquid surface water (determined as CHZ: Conventional Habitable Zone), but other more refined boundaries may be considered \citep{Kopparapu2013, Ramirez2019}. For example, the inner habitable boundary may be defined by critical fluxes, which cause runaway/moisture greenhouse effects \citep{Kasting1988} while the outer boundary may be constrained by the presence of carbon dioxide in the atmosphere as gas phase, avoiding its condensation \citep{Kasting1993}. The exoplanets within CHZs around active stars can be subject to high ionizing radiation fluxes including X-ray and Extreme Ultraviolet Emission (referred as to XUV (1-1200 $\mathrm{\AA}$) Emission),coronal mass ejections (CMEs) and associated stellar energetic particles (SEP) events  that can affect exoplanetary habitability conditions (\citealt{Airapetian2017a}; \citealt{Airapetian2019}).

Energetic stellar flare events associated with coronal mass ejections (CME) from magnetically active stars can contribute to the generation of stellar transient XUV emission and form high-energy particles accelerated in CME driven shocks (\citealt{Kumari2017}; \citealt{Gopalswamy2017}; \citealt{Airapetian2019}). These SEPs can penetrate into exoplanetary atmospheres, and cause chemical changes. These changes can be positive for the initiation of prebiotic chemistry in the planetary atmospheres or detrimental due to the destruction of a large fraction of ozone that transmits UVC (1000-2800  $\mathrm{\AA}$) and UVB (2800-3150  $\mathrm{\AA}$) emission to the exoplanetary surfaces
(\citealt{Airapetian2016}; \citealt{Airapetian2017b}; \citealt{Segura2010};  \citealt{Tilley2019}).

Our own Sun is known to exhibit extreme flare activity in the past including the so-called Carrington-class event \citep{Townsend2006}. Recent observations by the Kepler space telescope revealed that young solar-type stars generate much higher frequency of energetic flares (superflares), which could have been an important factor for habitability in the early history of our solar system and/or most extrasolar systems \citep{Maehara2012,Shibayama2013,Takahashi2016,Notsu2019, Airapetian2019}. Extreme surges of $^{14}\mathrm{C}$ were detected in the tree rings (\citealt{Miyake2012}; \citealt{Miyake2013}), which is considered strong evidence of the occurrence of superflares more than one magnitude stronger than the Carrington-class event \citep{Usoskin2013}. Effects of stellar activity from their host stars may also include periodic sterilizing doses of radiation via stellar superflare activity \citep{Lingam2017}. 
While the frequency and maximum energy of solar and stellar flares from planet hosts have not been well characterized, they may present a critical limiting factor on the development and persistence of life on terrestrial-type planets in our solar system (\citealt{Jakosky2015}; \citealt{Schrijver2015};\citealt{Kay2016}) as well as on Earth-sized exoplanets \citep{Atri2017}. 
Thus, a consistent approach to determine the habitable zone accounting for these factors is required. The characterization of these factors can be made using recently derived correlation between stellar flare frequency/intensity and starspot area, found from Kepler data, which may overcome the difficulty in prediction of flare impacted system \citep{Maehara2017}. 

Here, we present the first comprehensive impact evaluation system of expected ground level radiation doses in close-in terrestrial type exoplanets around M dwarfs including Proxima Centauri b (see Table \ref{tab:Table1}) in response to severe Solar Proton Events (SPEs). This study represents a realistic model of the surface dose evaluation for exoplanets with various possible atmospheric pressures and compositions. 
Section 1 presents the framework for evaluation of SPE particle fluence at the top of exoplanetary atmospheres. In Section 2 we discuss the application of the Particle and Heavy Ion Transport code System (PHITS) to a number of close-in exoplanets around M dwarfs. Section 3 discusses the ground dose for various exoplanetary systems and their consequences for the biological habitability of complex lifeforms. Section 4 describes the conclusions of the paper and future work.

\begin{longrotatetable}
\begin{deluxetable*}{lccc|cccccc|c}
\tablecaption{Basic parameters of target planets and their host stars, including their projected flare energy \label{tab:Table1}}
\tabletypesize{\footnotesize}
\tablehead{
\multicolumn{4}{c|}{Planet} & \multicolumn{7}{c}{Host star}
\\
\cline{1-4}
\cline{5-11}
\colhead{Exoplanet name} & \colhead{Radius} & \colhead{Size class} & 
\multicolumn{1}{c|}{Mass} & \colhead{Spectra} & \colhead{$T_{\mathrm{eff}}$} & 
\colhead{Radius} & \colhead{$P_{\mathrm{rot}}$} & \colhead{$A_{\mathrm{spot,p}}/(2\pi R_{\odot})$} & 
\multicolumn{2}{c}{Flare energy [erg]} 
\\
\cline{10-11}
\colhead{} & \colhead{[$R_{\mathrm{Earth}}$]} & \colhead{} & 
\multicolumn{1}{c|}{[$M_{\mathrm{Earth}}$]} & \colhead{Type} & \colhead{[K]} & 
\colhead{[$R_{\mathrm{\odot}}$]} & \colhead{[day]} & \colhead{} &
\multicolumn{1}{c|}{Annual} & \colhead{Spot Maximum} 
}
\startdata
GJ 699 b & 1.37 & super-Earth-size & 3.23 & M4V & 3278 & 0.18 & 140.0 & 0.0003 & 6.26E+31 & 1.15E+32 \\
Kepler-283 c & 1.82 & super-Earth-size & 4.59 & K5 & 4141 & 0.64 & 18.2 & 0.0021  & 4.93E+32 & 2.13E+33  \\ 
Kepler-1634 b & 3.19 & Neptune-size & 7.77 & G7 & 5637 & 0.82 & 19.8 & 0.0066  & 1.65E+33 & 1.18E+34  \\ 
Proxima Cen b & 1.07 & Earth-size & 1.27 & M5.5V & 3050 & 0.14 & 82.6 & 0.0040  & 9.7E+32 & 5.55E+33  \\ 
Ross-128 b & 1.10 & Earth-size & 1.40 & M4 & 3192 & 0.20 & 121.0 & 0.0002  & 4.72E+31 & 7.72E+31  \\ 
TRAPPIST-1 b & 1.09 & Earth-size & 0.86 & M8 & 2550 & 0.12 & 3.3 & 0.0012  & 2.70E+32 & 9.09E+32 \\ 
TRAPPIST-1 c & 1.06 & Earth-size & 1.38 & M8 & 2550 & 0.12 & 3.3 & 0.0012  & 2.70E+32 & 9.09E+32  \\ 
TRAPPIST-1 d & 0.77 & Earth-size & 0.41 & M8 & 2550 & 0.12 & 3.3 & 0.0012  & 2.70E+32 & 9.09E+32  \\ 
TRAPPIST-1 e & 0.92 & Earth-size & 0.64 & M8 & 2550 & 0.12 & 3.3 & 0.0012  & 2.70E+32 & 9.09E+32 \\ 
TRAPPIST-1 f & 1.05 & Earth-size & 0.67 & M8 & 2550 & 0.12 & 3.3 & 0.0012 & 2.70E+32 & 9.09E+32 \\ 
TRAPPIST-1 g & 1.13 & Earth-size & 1.34 & M8 & 2550 & 0.12 & 3.3 & 0.0012  & 2.70E+32 & 9.09E+32  \\ 
TRAPPIST-1 h & 0.76 & Earth-size & 0.36 & M8 & 2550 & 0.12 & 3.3 & 0.0012  & 2.70E+32 & 9.09E+32  \\ 
Sol d (Earth) & 1.00 & Earth-size & 1.00 & G2V & 5778 & 1.00 & 25.0 & 0.0030  & 7.20E+32 & 3.64E+33  \\ 
Sol e (Mars) & 0.53 & Mars-size & 0.11 & G2V & 5778 & 1.00 & 25.0 & 0.0030  & 7.2E+32 & 3.64E+33  \\ 
\enddata
\end{deluxetable*}
\end{longrotatetable}

\section{Method}
\subsection{Outline of Fluence Estimation for Top of Atmosphere (TOA) on Each Planet from Stellar Proton Events, and Definition of Maximum Flare Energy}

Our analysis is based on the application of stellar flare and starspot data derived mostly from the Kepler mission 
(\citealt{Maehara2015}; \citealt{Maehara2017}; \citealt{Notsu2013}; \citealt{Notsu2015a}; \citealt{Notsu2015b}; \citealt{Notsu2019}), in the ExoKyoto exoplanetary database (Yamashiki et al. 2019, in preparation). Our method utilizes starspot data derived from optical light curves to be used in parametric studies of the thickness of hypothetical exoplanetary atmospheres as the major attenuation factor of the incident radiation[see Table 1].
These data are used as input for the Particle and Heavy Ion Transport code System - PHITS \citep{Sato2018a} Monte-Carlo simulation model that is used for simulations of surface dose for terrestrial type exoplanets. 

\begin{figure}
\begin{center}
\includegraphics[width=1\textwidth]{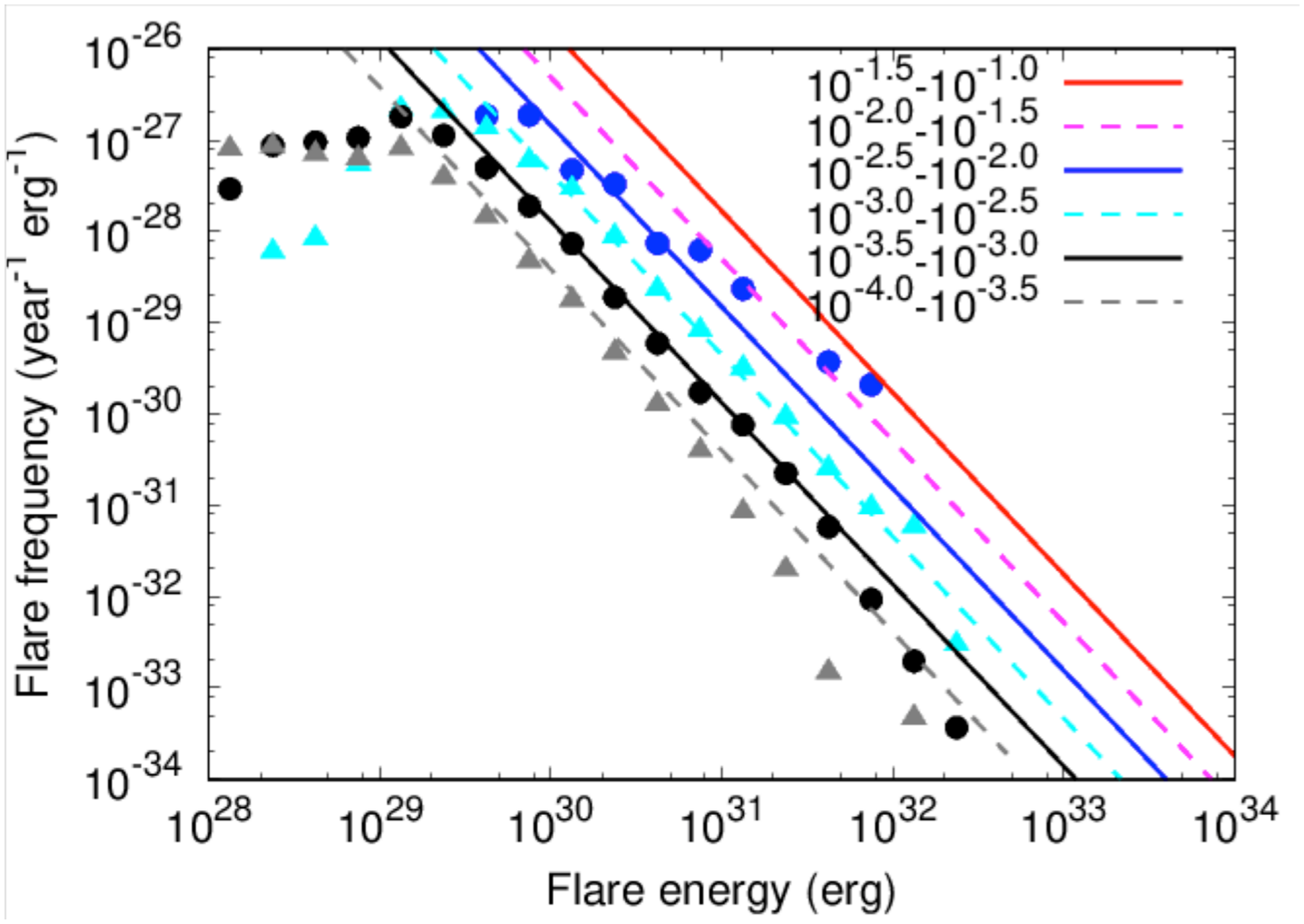}
\caption{Flare frequency vs flare energy for Solar Flares. The fraction of flare stars as a function of the rotation period. The solid line and dotted line represent the estimated scaling low calculated using equation (\ref{eq:dn_de}) as different star spot areas derived from \citet{Maehara2017}}
\label{fig:solarf-freq}
\end{center}
\end{figure}

\begin{figure}
\begin{center}
\includegraphics[width=1\textwidth]{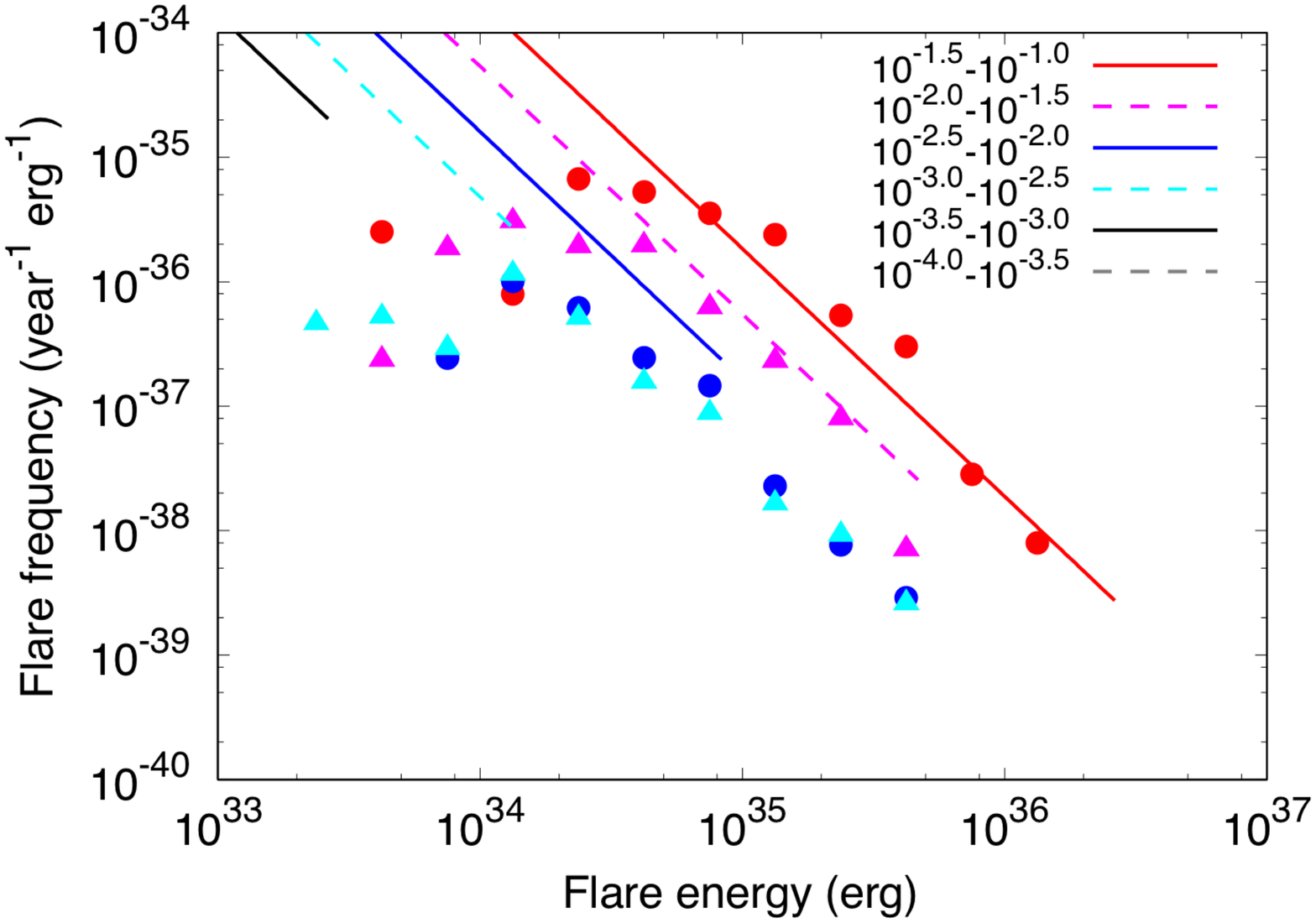}
\caption{Flare frequency vs flare energy for Solar Flares. The fraction of flare stars as a function of the rotation period. The solid line and dotted line represent the estimated scaling low calculated using equation (\ref{eq:dn_de}) as different star spot areas derived from \citet{Maehara2017}}
\label{fig:stellerf-freq}
\end{center}
\end{figure}

The following equations derive an assumed stellar flare magnitude from observed stellar spot size data. For the estimation of spot size, we used the same method in \citealt{Maehara2017}. Figures \ref{fig:solarf-freq} and \ref{fig:stellerf-freq} illustrate flare frequency vs flare energy for solar flares. The solid line and dotted line represent the estimated scaling low calculated using equation (\ref{eq:dn_de}) as a different starspot area derived from \citet{Maehara2017}.

Using the results of the above study, we derived the flare frequency distribution over its energy in the optical band as a function of the stellar spot size as follows:
\begin{equation}
\frac{dN}{dE} = C_{\mathrm{0}}\left[\mathrm{year}^{-1} \mathrm{erg}^{-1}\right] \left[ \frac{A_{\mathrm{spot}}}{10^{-2.75} A_{\mathrm{phot}}} \right]^{1.05} \left[ \frac{E_{\mathrm{flare}}}{10^{31} \mathrm{erg}} \right] ^ {-1.99},
\label{eq:dn_de}
\end{equation}
in which
$N$: Flare frequency (year $^{-1}$), 
$A_{\mathrm{spot}}$: Total area of starspots, 
$A_{\mathrm{phot}}$: Total visible area of the stellar surface,
$E_{\mathrm{flare}}$: Total expected stellar flare energy (erg), and
$C_{\mathrm{0}}$: Flare frequency constant (10$^{29.4}$).

Here we define $E_0 = 10 ^{31} \mathrm{erg}$ and set $N$ as 1, we then may determine Annual Maximum Flare energy as follows:

\begin{equation}
        E_{\mathrm{AMF}} = C_{\mathrm{0}}^{-\frac{1}{1+a}}  \left[ \frac{A_0}{A} \right]^{ \frac{b}{1+a}} E_0 ^ {\frac{a}{1+a}},
\label{eq:e_amf}
\end{equation}
in which 
$E_{\mathrm{AMF}}$: Annual Maximum Flare energy, as total expected stellar flare energy per year (erg  year$^{-1}$ ), $a=-1.99$, $b=1.05$.

The Spot Maximum Flare, maximum flare energy under a determined starspot area, can be illustrated as 
\begin{equation}
        E_{\mathrm{SMF}} = 7 \times 10 ^{32} (\mathrm{erg}) \left({\frac{f}{0.1}}\right)  \left( \frac{B_0}{10^{3}\mathrm{G}} \right)^{2} \left[{\frac{A_{\mathrm{spot}} / (2 \pi R_{\odot}^2)}{0.001}}\right]^{3/2},
\label{eq:e_smf}
\end{equation}
in which
$f$ : Fraction of magnetic energy that can be released as flare energy, 
$B_{\mathrm{0}}$ : Magnetic-field strength,
$E_{\mathrm{SMF}}$ : Spot Maximum Flare energy, as the theoretical maximum flare energy with a determined starspot area (erg), and
$R_{\odot}$: Solar radius ($7 \times 10^{10}$ cm ).

Possible Maximum Flare energy in this study was determined through the following (1) Evaluate maximum starspot coverage of the star through observation of stellar lightcurves. In this study, we observed a 20\% coverage of starspot on Proxima Centauri; accordingly we determined the maximum starspot coverage as 20\%. Then, (3) calculate maximum energy induced by the starspot area by \citet{Shibata2013}. 

The outline of the estimation method is as follows:

(STEP 1) Derive the magnitude and frequency of Stellar Proton Events from each star 
(1) by using direct observation of a stellar flare as a proxy of an SPE energy; and
(2) by applying the starspot area and/or rotational period correlation methodology. The conversion equation is presented and discussed in the next section. 
We use above information to extract representative starspot areas which can be applied to the conversion equations to flare energy expressed in the following section.  Accordingly we obtain (a) Annual Maximum flare (see equation(\ref{eq:e_amf}))  Spot Maximum flare (see equation(\ref{eq:e_smf})) (\citealt{Aschwanden2017}; \citealt{Shibata2013}), and (c) Possible Maximum Flare, calculated assuming that the target star surface is covered with starspots under the maximum percentage of observed starspot area (set as 20\% of half the spherical area). 
 
(STEP 2)
After the above procedure is completed for each star system, the possible quantitative exposures are assumed by the following procedure:
(4) Estimating the fluence of each Stellar Proton Event at the TOA using the equation (\ref{eq:e_spe_earth}) .

As for the atmospheric compositions of exoplanets, three-types of atmospheres for typical extrasolar planets are considered (explained in detail in the following section). For those typical atmospheric compositions, the potential doses for life on extrasolar planets are determined through the following procedure.

(STEP 3)
(5) Calculate the possible dose rate from the Monte-Carlo simulations using Particle and Heavy Ion Transport code System PHITS  \citep{Sato2018a}  for three typical atmospheric compositions as extrasolar planetary atmospheres,
(6) Normalize the dose by determining the Earth equivalent ratio, which was previously normalized by using
(6a) The Carrington-class event, assuming that the event has X45 class, or by
(6b) The deepest observed flare event GLE43 which occurred in 1989, as X13 class \citep{Xapsos2000}. 
(7) Calculate conversion coefficients for each exoplanet by comparing the values calculated in (4) and (6).
(8) Convert the reference dose value calculated in (6) into each extrasolar planet case using conversion coefficients.


\subsection{Monte Carlo Simulation for Air-Shower using PHITS}

When high-energy SEPs precipitate into the planetary atmosphere, they induce extensive air-shower (EAS) by producing various secondary particles, such as neutrons and muons. We conducted a three-dimensional EAS simulation by using the Particle and Heavy Ion Transport code System, PHITS \citep{Sato2018a}, which is a general-purpose Monte Carlo code for analyzing the propagation of radiation in any materials. PHITS version 2.88 with the recommended setting for cosmic-ray transport simulation \citep{Sato2014} was used in this survey. In our simulations we assume the size and the mass of the modeled planets to be the same as that of the Earth.  

\subsection{Chemical Composition of Exoplanetary Atmospheres} 

The impact of stellar proton events on a planet depends upon its atmospheric composition. We consider three: Earth-like (N$_2$+O$_2$ rich), Mars-like or Venus-like (CO$_2$ rich) and an young Earth-sized or super-Earth's with a primary H$_2$ rich atmospheres. We assume that the Earth-type atmosphere is the standard land-ocean planetary atmosphere composed mostly of Nitrogen and Oxygen (N$_2$+O$_2$). A Venusian-like atmosphere is represented as a CO$_2$ -rich atmosphere resulted from the runaway greenhouse effect and subsequent outgassing of CO$_2$ from carbonates. The Martian-like atmosphere is an example of a low gravity low pressure planetary CO$_2$ -rich atmosphere that has experienced severe atmospheric escape driven by strong stellar ionizing radiation flux. We also model a young Earth-sized H$_2$ rich atmosphere, because such an atmosphere is assumed for large super-Earth planets, whose gravitational pull might be sufficiently large to retain substantial atmospheric H$_2$. Hydrogen rich atmospheres of Earth-sized exoplanets can be formed due to capture of hydrogen from protoplanetary atmospheres and/or during accretion period (\citealt{Elkins2008}; \citealt{Lammer2018}). Thus, here we refer to young Earth-sized exoplanets. 

The composition of the atmosphere, for the above three typical atmospheric types, were set to 78\% nitrogen, 21\% oxygen and 1\% argon for the Earth-like (N$_2$+O$_2$) atmosphere, for the Martian/Venusian-type (CO$_2$), and 100\% hydrogen for the young-Earth-type (H$_2$).  During the Monte-Carlo simulation using PHITS, we assume the composition of the planet interior to be covered with sufficient liquid water for the Terrestrial-type, while the same gas was continuously filled in the planet interior for the other cases. In our simulation of all model atmosphere (young Earth-type(H$_2$), Earth-like (N$_2$+O$_2$), Martian and Venusian-type (CO$_2$)) cases we assume the exoplanet radius and mass to be 1 R$_{Earth}$ and 1 M$_{Earth}$, respectively. Numerical simulation of super-Earths will be performed in the upcoming studies.

\subsection{Event Integrated Spectra of Extreme SPEs}

We assume that stellar accelerated protons are isotropically distributed in space as they precipitate into the atmospheres of the modeled planets and have two different energy spectra represented by the SPE spectra derived for the Carrington-class event in 1859 \citep{Townsend2006} and the 43rd ground level enhancement (GLE) in 1989 \citep{Xapsos2000}, respectively. 

The Carrington-class event is considered to be the largest eruptive event recorded in modern human history.
However, according to \citet{Smart2006}, proton energy spectra associated with the Carrington-class event was rather soft in comparison with other solar flares that produce GLEs.  Thus, the radiation dose at the ground level during the event is expected to be not significantly high, because only a small fraction of the high-energy protons (with energies over 3 GeV for an 1-bar atmosphere) and their secondary particles can penetrate into the deep atmosphere. 

To estimate the maximum impact on the ground level, we therefore calculated the radiation dose during the solar flare in association with a harder proton spectrum, GLE43, which is one of the most significant GLE that has occurred after satellite observations were started in the late 20th century. It should be noted that GLE 43 was selected as a typical SPE associated with a hard proton spectrum to estimate the maximum impact of SPE exposure at deeper locations in the atmosphere, though its flare class was not extremely high (X13). GLE 5 (23 Feb 1956) type spectra were also considered as a relevant event for the survey. 

\begin{figure}
\begin{center}
\includegraphics[width=1\textwidth]{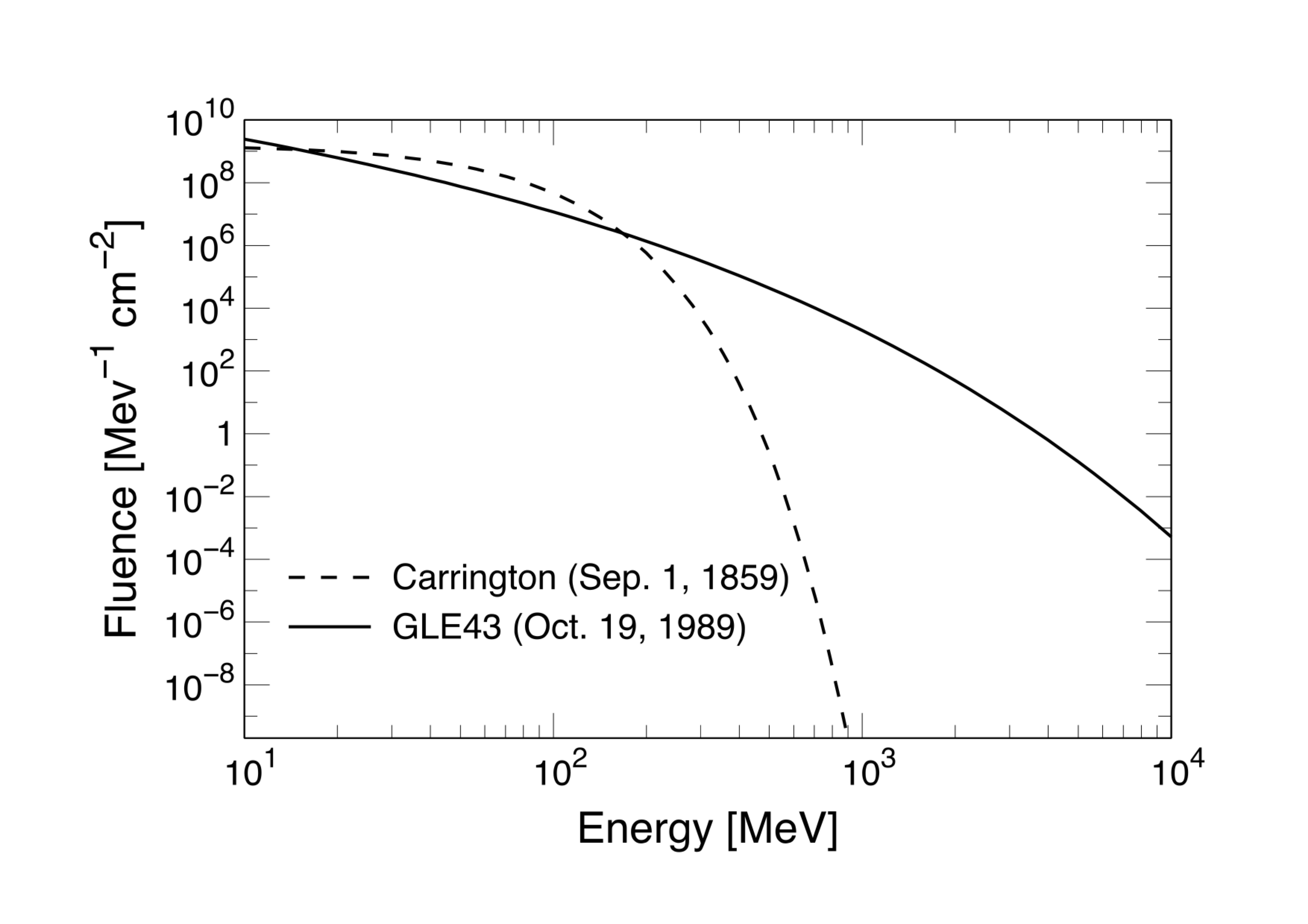}
\caption{Event integrated spectra, of the GLE43 that occurred on 19 October 1989 (solid) and the Carrington Flare that occurred on 1 September 1859 (dotted) solar proton events on Earth, based on parameters obtained from references.}
\label{fig:ei-spectra}
\end{center}
\end{figure}

Figure \ref{fig:ei-spectra} illustrates event-integrated spectra of the GLE43 that occurred on 19 October 1989 (solid) and the Carrington Flare that occurred on 1 September 1859 (dotted) solar proton events, based on parameters obtained from references.

\subsection{The Influence of the Planetary Magnetic Field } 

We have simulated four scenarios of exoplanetary dipole magnetic moments: (i) B = 0 (unmagnetized planet), (ii) $0.1 \times B_{\mathrm{Earth}}$, (iii) $1 \times  B_{\mathrm{Earth}}$ (Earth-likle magnetic moment),  and (iv) $10 \times B_{\mathrm{Earth}}$. 
The impact of the planetary magnetic field on the surface dose was modeled via the magnetospheric filter functions for the above 4 different magnetic moments, 0, 0.1, 1, and 10 $ B_{\mathrm{Earth}}$, evaluated by \citet{Grie2015}. 

The fluence of protons, neutrons, positive and negative muons, electrons, positrons, and photons were scored as a function of the atmospheric depth. They were then converted to the absorbed dose in Gy and the effective dose in Sv, using the stopping power and the fluence to the dose conversion coefficients for the isotropic irradiation \citep{ICRP2010}, respectively. It should be mentioned that the effective dose is defined as only used for the purpose of radiological protection. However we evaluated it for discussing the possible exposure effects on human-like lifeforms because there is no alternative quantity that can be used for this discussion.
More detailed descriptions on the simulation procedures as well as their verification results for the solar energetic particle and galactic cosmic-ray simulation in the terrestrial atmosphere were given in our previous papers (\citealt{Sato2015}; \citealt{Sato2018b}).

The impacts of all components produced by cosmic-ray interactions with the different atmospheric types in different layers were also individually evaluated and finally integrated to produce a final ground level dose value for each simulated scenario. By examining all these different parameters together (atmospheric composition, geomagnetic field strength and simulated cosmic ray interactions), we have evaluated the atmospheric barrier needed for life on each of the target planets to survive a stellar flare event. This approach assumes that the potential life is as similarly radiation tolerant as that present on Earth.

\subsection{The Maximum Stellar Flare Energy}
Our goal is to study the effect of high ionizing particle fluxes caused by stellar activity on habitability of close-in Earth-sized and super Earth exoplanets located within habitable zones. CHZs around low luminosity M dwarfs are located within 0.05 AU that suggests that many of them orbit their host stars within sub Alfvenic distance and are subject to direct irradiation via high particle fluxes. To study the resulted surface dose we selected four exoplanets around active M dwarfs, one exoplanet around K dwarfs with detected superflare, and one exoplanet around G dwarf with higher stellar activity than our sun.
 We selected the target stars for this survey according to the following procedure (1) Select host star with exoplanet in habitable zone with direct superflare observation through Kepler observation (Kepler-283) (2) Select Kepler stars whose flare frequency and magnitude can be estimated from their activities (Kepler-1634) and (3) Select well-documented host star for well-documented exoplanets (GJ699 (Barnards Star), Proxima Centauri, Ross-128, TRAPPIST-I). Stellar activities for all stars are estimated using their light curves.

\citet{Shibata2013} estimated the maximum value (upper limit) of flare energy, which is determined by the starspot area and magnetic field strength. We used this methodology to calculate the theoretical maximum flare energy for six host stars using their starspot areas: $ 1.15 \times 10 ^{32} $ erg for GJ 699 (Barnard's Star), $ 2.13 \times 10 ^{33} $ erg for Kepler-283, $1.18 \times 10 ^{34} $ erg for Kepler-1634, $5.55 \times 10^{33} $erg for Proxima Centauri, $ 7.72 \times 10 ^{31}$ erg for Ross 128, and $ 9.09 \times 10 ^{32} $ erg for TRAPPIST-1.
 
In this method, the current observed starspot area in each star restricts the maximum flare energy. However, it is unclear whether the observed period represents the maximum or minimum activity of the star. Accordingly, we also evaluated the potential maximum energy of the stellar flare by the following method.

For those stars whose stellar temperature is above 4000 K: we estimated maximum flare energy based on the relationship between Kepler stars by comparing their maximum observed energy and stellar temperature as well as its associated radius (H. Maehara, private communication).

For those stars whose stellar temperature is below 4000 K: we assumed, in the extreme situation, that 20 \% of the stellar surface is covered by starspot. Considering the extreme condition, we calculated the maximum energy using \citet{Shibata2013}.

By introducing flare energy as input for considerable maximum energy of the superflares for their planetary systems, we may theoretically calculate the possible maximum dose for their host planets.

\section{Results}
\subsection{Validation for Normal Dose }

\begin{figure}
\begin{center}
\includegraphics[width=1\textwidth]{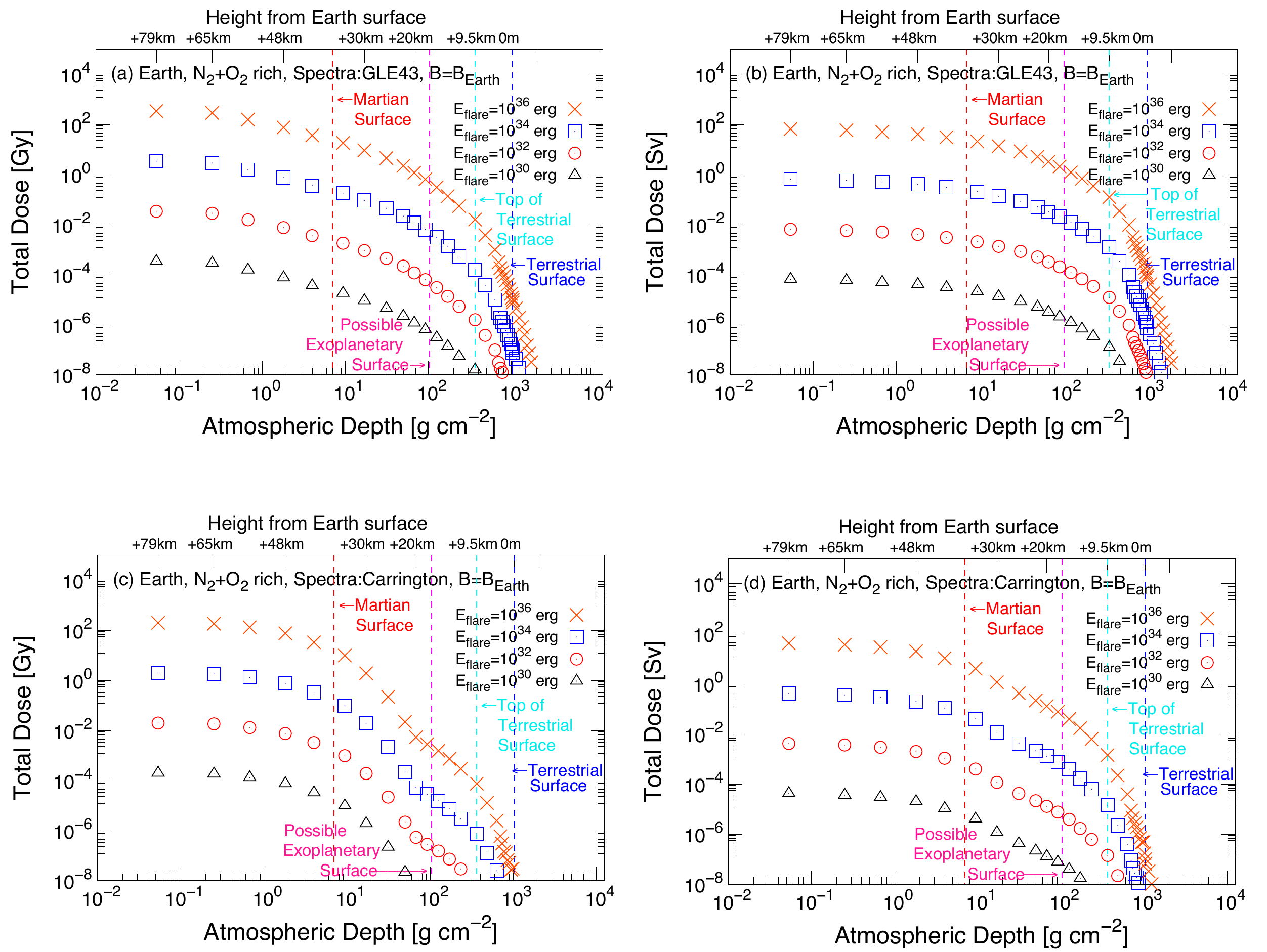}
\caption{Vertical profile of radiation dose on Earth for normalized flares, caused by hard proton spectrum (imitating GLE 43) (a)(b) and soft spectrum (imitating Carrington by Townsend) (c)(d) penetrating N$_2$+O$_2$ rich (terrestrial type) atmosphere for Earth with $10^{30}$  erg (black triangle), $10^{32}$ erg (red circle), $10^{34}$ erg (blue square) and $10^{36}$ erg (red cross) in Gray (Gy) (a) (c) and Sievert (Sv) (b) (d). The vertical legend shows the following four typical atmospheric depth reference layers: Martian Surface Atmospheric Pressure equivalent to 9 g/cm$^2$, Terrestrial Minimum Atmospheric Pressure, observed at the summit of the Himalayas equivalent to 365 g/cm$^2$ in this study, and (Earth's) Ground Level Atmospheric Pressure, equivalent to 1037 g/cm$^2$.  Possible Exoplanetary Surface was estimated as 1/10 of terrestrial surface equivalent to 103.7 g/cm$^2$. Note that the value is not identical to the real observation data but at the nearest value empolyed in the Monte-Carlo numerical simulation using PHITS.
}
\label{fig:Earth-Norm-Flare}
\end{center}
\end{figure}

Figure \ref{fig:Earth-Norm-Flare} shows the vertical profile of radiation dose on Earth and Mars caused by SPEs with the hard proton spectrum (imitating GLE 43) (a)(b) and soft spectrum (imitating Carrington) (c)(d) penetrating N$_2$+O$_2$  rich (terrestrial-type) atmosphere Earth with $10^{30}$ erg (black triangle), $10^{32}$ erg (red circle), $10^{34}$ erg (blue square) and $10^{36}$  erg (red cross) in Gray (Gy) (a) (c) and Sievert (Sv) (b) (d). This figure shows that the radiation dose at the tropopose (around 170 g/cm$^2$ atmospheric depth) becomes 0.5 Millisievert which agrees mostly with the aerial observation, when the solar flare energy is scaled to $E_0 = 10^{32}$. Note that this normalization has been made for an idealized series of flares, considering the horizontal angle of the SPE injection as 90 degrees, in other words, the probability of reaching Earth is 1/4.

\begin{figure}
\begin{center}
\includegraphics[width=1\textwidth]{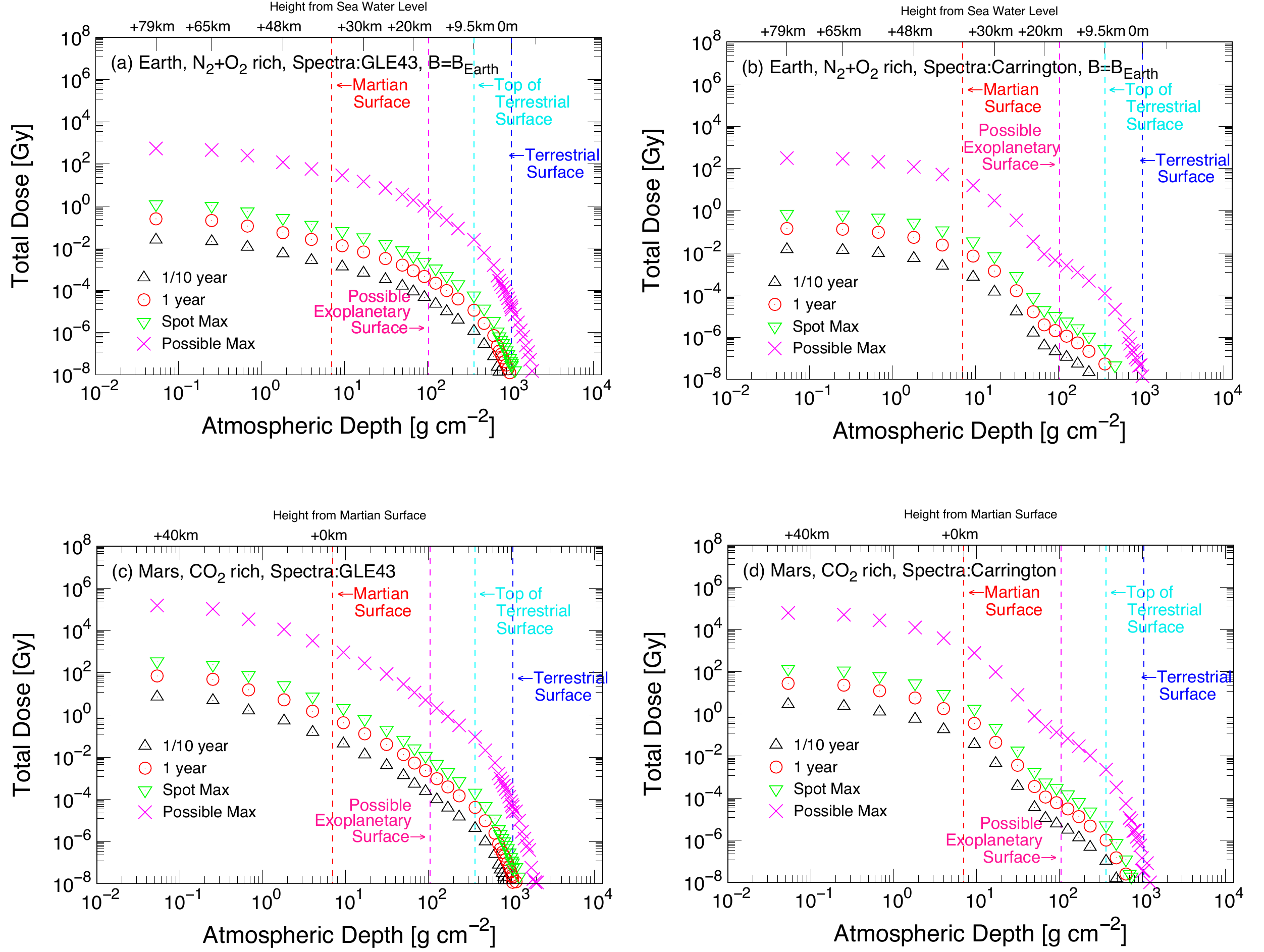}
\caption{Vertical profile of radiation dose(Gy)on Earth and Mars for possible flares on several different scales, caused by hard proton spectrum (imitating GLE 43) (a)(c) and soft spectrum (imitating Carrington) (b)(d) penetrating N$_2$+O$_2$ rich (terrestrial type) atmosphere for Earth (a)(b) and CO$_2$ rich (Martian type) atmosphere for Mars(c)(d) with flares every 1/10 year (36 days, black triangle), one year (red circle), spot maximum (green triangle), possible max (red cross). Martian Surface Atmospheric Pressure, equivalent to 9 g/cm$^2$; Terrestrial Minimum Atmospheric Pressure, observed at the summit of the Himalayas equivalent to 365 g/cm$^2$; (Earth's) Ground Level Atmospheric Pressure, equivalent to 1037 g/cm$^2$;  Possible Exoplanetary Surface, 1/10 of terrestrial surface equivalent to 103.7 g/cm$^2$.}
\label{fig:earth-mars-flare-Gy}
\end{center}
\end{figure}

\begin{figure}
\begin{center}
\includegraphics[width=1\textwidth]{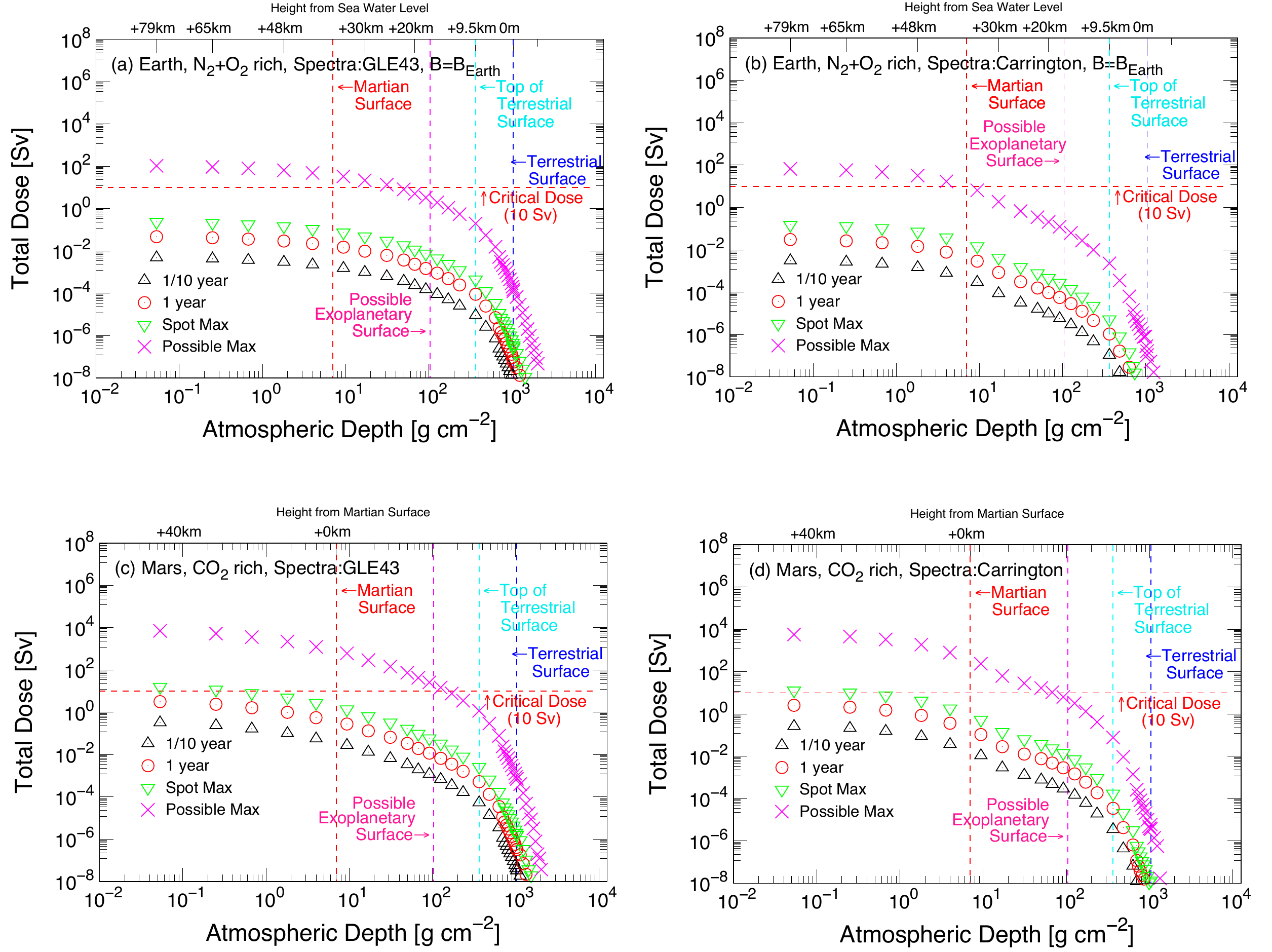}
\caption{Vertical profile of radiation dose(Sv)on Earth and Mars for possible flares on several different scales, caused by hard proton spectrum (imitating GLE 43) (a)(c) and soft spectrum (imitating Carrington) (b)(d) penetrating N$_2$+O$_2$ rich (terrestrial type) atmosphere for Earth (a)(b) and CO$_2$ rich (Martian type) atmosphere for Mars(c)(d) with flares  every 1/10 year (36 days, black triangle), one year (red circle), spot maximum (green triangle), possible max (red cross). Martian Surface Atmospheric Pressure, equivalent to 9 g/cm$^2$; Terrestrial Minimum Atmospheric Pressure, observed at the summit of the Himalayas equivalent to 365 g/cm$^2$; (Earth's) Ground Level　Atmospheric Pressure, equivalent to 1037 g/cm$^2$;  Possible Exoplanetary Surface, 1/10 of terrestrial surface equivalent to 103.7 g/cm$^2$. }
\label{fig:earth-mars-flare-Sv}
\end{center}
\end{figure}

Figures \ref{fig:earth-mars-flare-Gy} and \ref{fig:earth-mars-flare-Sv} show vertical profile of radiation dose in Gray (\ref{fig:earth-mars-flare-Gy}) and in Sievert(\ref{fig:earth-mars-flare-Sv}) on Earth and Mars for possible flares on several different scales, caused by hard proton spectrum (imitating GLE 43) (a)(c) and soft spectrum (imitating Carrington reproduced by \citep{Townsend2006}) (b)(d) penetrating N$_2$+O$_2$ rich (terrestrial-type) atmosphere for Earth (a)(b) and CO$_2$ rich (Martian type) atmosphere for Mars(c)(d) with flares every 1/10 year (36 days, corresponding 7.2 $\times 10^{31}$ erg), one year (corresponding 7.2 $\times 10^{32}$ erg), Spot Maximum flare (corresponding 3.6 $\times 10^{33}$ ergs) and Possible Maximum flare (corresponding 1.6 $\times 10^{36}$ erg). In these scenarios, the Spot Maximum flare is the maximum possible flare to be observed within decades in the target stellar system (in this case our solar system) estimated based on starspot area of the target star. According to the calculation shown in these figures, the Solar Proton Events under the above scenarios do not induce a critical dose at ground level when we have sufficient atmospheric depth such as Earth, even under Possible Maximum flare (1.6 $\times 10^{36}$ erg) scenario, whereas it becomes a nearly critical dose on the  Martian surface with thinner atmospheric depth when the Spot Maximum flare (3.6 $\times 10^{33}$ ergs) event occurs. 

\subsection{Estimated Dose on Exoplanetary Surface Under Different Flare Scenarios }
The estimated doses under the Annual Maximum flare and under the Spot Maximum flare are shown in Tables \ref{tab:Table2} and \ref{tab:Table3}, respectively. The estimated doses at the Top of Atmosphere (TOA) on GJ 699 b, Proxima Cen b, Ross-128 b and TRAPPIST-1 e under the Spot Maximum Flare \citep{Shibata2013}  becomes 1.60 $\times 10^2 $Gy (7.25 Sv), 5.36 $\times 10^5 $Gy (2.43 $\times 10^4 $Sv) ,  7.13 $\times 10^3$ Gy (3.23 $\times 10^2$ Sv), and 2.60 $\times 10^5$ Gy (1.18 $\times 10^4$ Sv), respectively.

\begin{table}[ht!]

\centering
\renewcommand{\thetable}{\arabic{table}}
\caption{Estimated dose under projected flare event - Annual Maximum Flare}
\label{tab:Table2}
\begin{tabular}{|l|c|c|c|c|c|c|c|c|c|c|c|c|c|c|c|c|c|}
\tablewidth{0pt}
\hline 
\hline
Exoplanet 
& TOA
& TOA
& MS
& MS
& TM
& TM
& GL
& GL \\
Name & [Gy] & [Sv] & [Gy] & [Sv] & [Gy] & [Sv] & [Gy] & [Sv] \\
 & [1] & [2] & [3] & [4] 
 & [5] & [6] & [7] & [8] \\

\hline
GJ 699 b & 8.72E+01 & 3.95E+00 & 5.20E-01 & 3.46E-01 & 5.82E-05 & 5.91E-04 & 2.31E-08 & 2.59E-07 \\
Kepler-283 c & 	9.64E+02 & 4.36E+01 & 5.75E+00 & 3.83E+00 & 6.44E-04 & 6.54E-03 & 	2.56E-07 & 2.86E-06 \\
Kepler-1634 b & 3.84E+02 & 1.74E+01 & 2.29E+00 & 1.53E+00 & 2.56E-04 & 2.61E-03 & 1.02E-07 & 1.14E-06 \\
Proxima Centauri b & 9.37E+04 & 4.24E+03 & 5.60E+02 & 3.72E+02 & 6.26E-02 & 6.36E-01 & 2.49E-05 & 2.79E-04 \\
Ross 128 b & 4.36E+03 & 1.97E+02 & 2.60E+01 & 1.73E+01 & 2.91E-03 & 2.96E-02 & 1.16E-06 & 1.30E-05 \\
TRAPPIST-1 b & 4.97E+05 & 2.25E+04 & 2.97E+03 & 1.97E+03 & 3.32E-01 & 3.37E+00 & 1.32E-04 & 1.48E-03 \\
TRAPPIST-1 c & 2.65E+05 & 1.20E+04 & 1.58E+03 & 1.05E+03 & 1.77E-01 & 1.80E+00 & 7.04E-05 & 7.88E-04 \\
TRAPPIST-1 d & 1.33E+05 & 6.04E+03 & 7.97E+02 & 5.30E+02 & 8.91E-02 & 9.06E-01 & 3.54E-05 & 3.97E-04\\
TRAPPIST-1 e & 7.73E+04 & 3.50E+03 & 4.61E+02 & 3.07E+02 & 5.16E-02 & 5.25E-01 & 2.05E-05 & 2.30E-04 \\
TRAPPIST-1 f & 4.46E+04 & 2.02E+03 & 2.66E+02 & 1.77E+02 & 2.98E-02 & 3.02E-01 & 1.18E-05 & 1.32E-04 \\
TRAPPIST-1 g & 3.02E+04 & 1.37E+03 & 1.80E+02 & 1.20E+02 & 2.01E-02 & 2.05E-01 & 8.00E-06 & 8.96E-05 \\
TRAPPIST-1 h & 1.55E+04 & 7.00E+02 & 9.22E+01 & 6.14E+01 & 1.03E-02 & 1.05E-01 & 4.10E-06 & 4.59E-05 \\
Sol d (Earth) & 1.64E+02 & 7.40E+00 & 9.76E-01 & 6.50E-01 & 1.09E-04 & 1.11E-03 & 4.34E-08 & 4.86E-07 \\
Sol e (Mars) & 7.05E+01 & 3.19E+00 & 4.21E-01 & 2.80E-01 & 4.71E-05 & 4.78E-04 & 1.87E-08 & 2.09E-07 \\

\hline
\multicolumn{9}{l}{[1] Estimated Dose [Gy] by Annual Maximum Flare at TOA}\\
\multicolumn{9}{l}{[2] Estimated Dose [Sv] by Annual Maximum Flare at TOA}\\
\multicolumn{9}{l}{[3] Estimated Dose [Gy] by Annual Maximum Flare at MS}\\
\multicolumn{9}{l}{[4] Estimated Dose [Sv] by Annual Maximum Flare at MS}\\
\multicolumn{9}{l}{[5] Estimated Dose [Gy] by Annual Maximum Flare at TM }\\
\multicolumn{9}{l}{[6] Estimated Dose [Sv] by Annual Maximum Flare at TM}\\
\multicolumn{9}{l}{[7] Estimated Dose [Gy] by Annual Maximum Flare at GL}\\
\multicolumn{9}{l}{[8] Estimated Dose [Sv] by Annual Maximum Flare at GL} \\
\multicolumn{9}{l}{TOA - Top of Atmosphere ($\approx$ 0 g/cm$^2$)}\\ 
\multicolumn{9}{l}{MS - Martian Surface Atmospheric Pressure (9 g/cm$^2$)}\\
\multicolumn{9}{l}{TM - Terrestrial Minimum Atmospheric Pressure (365 g/cm$^2$)}\\
\multicolumn{9}{l}{GL - (Earth's) Ground Level Atmospheric Pressure (1037 g/cm$^2$)}

\\

\end{tabular}

\end{table}

\begin{table}[ht!]

\centering
\renewcommand{\thetable}{\arabic{table}}
\caption{Estimated dose under projected flare event - Spot Maximum Flare}
\label{tab:Table3}
\begin{tabular}{|l|c|c|c|c|c|c|c|c|c|c|c|c|c|c|c|c|c|}
\tablewidth{0pt}
\hline 
\hline
Exoplanet 
& TOA
& TOA
& MS
& MS
& TM
& TM
& GL
& GL
\\
Name & [Gy] & [Sv] & [Gy] & [Sv] & [Gy] & [Sv] & [Gy] & [Sv]\\
& [9] & [10] & [11] & [12] & [13] & [14] & [15] & [16] \\

\hline
GJ 699 b & 1.60E+02 & 7.25E+00 & 9.56E-01 & 6.37E-01 & 1.07E-04 & 1.09E-03 & 4.25E-08 & 4.76E-07\\
Kepler-283 c & 4.16E+03 & 1.89E+02 & 2.49E+01 & 1.65E+01 & 	2.78E-03 & 2.83E-02 & 1.11E-06 & 1.24E-05\\
Kepler-1634 b & 2.74E+03 & 1.24E+02 & 1.64E+01 & 1.09E+01 & 1.83E-03 & 1.86E-02 & 7.27E-07 & 8.14E-06\\
Proxima Centauri b & 5.36E+05 & 2.43E+04 & 3.20E+03 & 2.13E+03 & 3.58E-01 & 3.64E+00 & 1.42E-04 & 1.59E-03\\
Ross 128 b & 7.13E+03 & 3.23E+02 & 4.26E+01 & 2.83E+01 & 4.77E-03 & 4.84E-02 & 1.89E-06 & 2.12E-05\\
TRAPPIST-1 b & 1.67E+06 & 7.58E+04 & 9.99E+03 & 6.65E+03 & 1.12E+00 & 1.14E+01 & 4.44E-04 & 4.97E-03\\
TRAPPIST-1 c & 8.93E+05 & 4.04E+04 & 5.33E+03 & 3.55E+03 & 5.96E-01 & 6.06E+00 & 2.37E-04 & 2.65E-03\\
TRAPPIST-1 d & 4.49E+05 & 2.03E+04 & 2.68E+03 & 1.79E+03 & 3.00E-01 & 3.05E+00 & 1.19E-04 & 1.34E-03\\
TRAPPIST-1 e & 2.60E+05 & 1.18E+04 & 1.55E+03 & 1.03E+03 & 1.74E-01 & 1.77E+00 & 6.91E-05 & 7.74E-04\\
TRAPPIST-1 f & 1.50E+05 & 6.79E+03 & 8.96E+02 & 5.96E+02 & 1.00E-01 & 1.02E+00 & 3.98E-05 & 4.46E-04\\
TRAPPIST-1 g & 1.02E+05 & 4.60E+03 & 6.06E+02 & 4.04E+02 & 6.78E-02 & 6.89E-01 & 2.69E-05 & 3.02E-04\\
TRAPPIST-1 h & 5.20E+04 & 2.36E+03 & 3.11E+02 & 2.07E+02 & 3.48E-02 & 3.53E-01 & 1.38E-05 & 1.55E-04\\
Sol d (Earth) & 8.27E+02 & 3.74E+01 & 4.93E+00 & 3.29E+00 & 5.52E-04 & 5.61E-03 & 2.19E-07 & 2.46E-06\\
Sol e (Mars) & 3.56E+02 & 1.61E+01 & 2.13E+00 & 1.42E+00 & 2.38E-04 & 2.42E-03 & 9.45E-08 & 1.06E-06\\

\hline
\multicolumn{9}{l}{[1] Estimated Dose [Gy] by Annual Maximum Flare at TOA}\\
\multicolumn{9}{l}{[2] Estimated Dose [Sv] by Annual Maximum Flare at TOA}\\
\multicolumn{9}{l}{[3] Estimated Dose [Gy] by Annual Maximum Flare at MS}\\
\multicolumn{9}{l}{[4] Estimated Dose [Sv] by Annual Maximum Flare at MS}\\
\multicolumn{9}{l}{[5] Estimated Dose [Gy] by Annual Maximum Flare at TM }\\
\multicolumn{9}{l}{[6] Estimated Dose [Sv] by Annual Maximum Flare at TM}\\
\multicolumn{9}{l}{[7] Estimated Dose [Gy] by Annual Maximum Flare at GL}\\
\multicolumn{9}{l}{[8] Estimated Dose [Sv] by Annual Maximum Flare at GL} \\
\multicolumn{9}{l}{TOA - Top of Atmosphere ($\approx$ 0 g/cm$^2$)}\\ 
\multicolumn{9}{l}{MS - Martian Surface Atmospheric Pressure (9 g/cm$^2$)}\\
\multicolumn{9}{l}{TM - Terrestrial Minimum Atmospheric Pressure (365 g/cm$^2$)}\\
\multicolumn{9}{l}{GL - (Earth's) Ground Level Atmospheric Pressure (1037 g/cm$^2$)}

\end{tabular}

\end{table}

\begin{table}[ht!]
\centering
\footnotesize
\renewcommand{\thetable}{\arabic{table}}
\caption{UV energy from annual maximum flare at top of atmosphere (TOA) in each planet}
\label{tab:Table4}
\begin{tabular}{|l|c|c|c|c|c|c|c|}
\tablewidth{0pt}
\hline 
\hline
Exoplanet 
& $E^{\rm{flare}}_{\rm{UV}}$ 
& $\frac{E^{\rm{flare}}_{\rm{UV}}}{E^{\rm{flare}}_{\rm{UV, Earth}}}$ 
& $\frac{E^{\rm{flare}}_{\rm{UV}}}{E^{\rm{flux}}_{\rm{UV, Earth}}}$ 
& $E^{\rm{flare}}_{\rm{XUV}}$
& $\frac{E^{\rm{flare}}_{\rm{XUV}}}{E^{\rm{flare}}_{\rm{XUV, Earth}}}$ 
& $\frac{E^{\rm{flare}}_{\rm{XUV}}}{E^{\rm{flux}}_{\rm{XUV, Earth}}}$ 
& $E^{\rm{normal}}_{\rm{XUV}}$ 
\\
Name & [J m$^{-2}$] & &  [\%] & [J m$^{-2}$]& & [\%] & [J m$^{-2}$] \\
 & [1] & [2] & [3] & [4] 
 & [5] & [6] & [7] \\
\hline
GJ 699 b & 8.74E+03 & 7.27E+04 & 2.59E$-$06 & 4.37E+03 &7.27E+04 & 2.51E$-$02 & 2.35E+04 \\
Kepler-283 c & 4.43E+02 & 3.69E+03 & 1.31E$-$07 & 2.22E+02 & 3.69E+03 & 1.27E$-$03 & 1.19E+06 \\ 
Kepler-1634 b & 4.60E+04 & 3.83E+05 & 1.37E$-$05 & 2.30E+04 & 3.83E+05 & 1.32E$-$01 & 6.67E+05  \\ 
Proxima Cen b & 1.57E+04 & 1.31E+05 & 4.66E$-$06 & 7.86E+03 & 1.31E+05 & 4.51E$-$02 & 1.33E+07  \\ 
Ross-128 b & 8.75E+04 & 7.28E+05 & 2.60E$-$05 & 4.38E+04 & 7.28E+05 & 2.51E$-$01 & 1.38E+06  \\ 
TRAPPIST-1 b & 5.30E+04 & 4.41E+05 & 1.57E$-$05 & 2.65E+04 & 4.41E+05 & 1.52E$-$01 & 5.99E+07 \\ 
TRAPPIST-1 c & 2.82E+04 & 2.35E+05 & 8.39E$-$06 & 1.41E+04 & 2.35E+05 & 8.11E$-$02 & 3.20E+07   \\ 
TRAPPIST-1 d & 1.42E+04 & 1.18E+05 & 4.22E$-$06 & 7.12E+03 & 1.18E+05 & 4.08E$-$02 & 1.61E+07   \\ 
TRAPPIST-1 e & 8.24E+03 & 6.86E+04 & 2.45E$-$06 & 4.12E+03 & 6.86E+04 & 2.37E$-$02 & 9.32E+06  \\ 
TRAPPIST-1 f & 4.75E+03 & 3.96E+04 & 1.41E$-$06 & 2.38E+03 & 3.96E+04 & 1.36E$-$02 & 5.38E+06   \\ 
TRAPPIST-1 g & 3.22E+03 & 2.68E+04 & 9.54E$-$07 & 1.61E+03 & 2.68E+04 & 9.23E$-$03 & 3.64E+06   \\ 
TRAPPIST-1 h & 1.65E+03 & 1.37E+04 & 4.89E$-$07 & 8.24E+02 & 1.37E+04 & 4.73E$-$03 & 1.86E+06   \\ 
Sol d (Earth) & 3.70E$-$01 & 3.08E+00 & 1.10E$-$10 & 1.85E$-$01 & 3.08E+00 & 1.06E$-$06 & 1.74E+05  \\ 
Sol e (Mars) & 1.59E$-$01 & 1.33E+00 & 4.72E$-$11 & 7.96E$-$02 & 1.33E+00 & 4.57E$-$07 & 7.51E+04  \\ 
\hline
\hline
Exoplanet 
& $E^{\rm{normal}}_{\rm{UV}} $ 
& $E^{\rm{normal}}_{\rm{Visible}}$ 
& $E^{\rm{normal}}_{\rm{IR}}$ 
& $E^{\rm{flare+quiescent}}_{\rm{XUV}}$ 
& $\frac{E^{\rm{flare+quiescent}}_{\rm{XUV}}}{E^{\rm{flare+quiescent}}_{\rm{XUV, Earth}}}$ 
& $E^{\rm{flare+quiescent}}_{\rm{UV}}$ 
& $\frac{E^{\rm{flare+quiescent}}_{\rm{UV}}}{E^{\rm{flare+quiescent}}_{\rm{UV, Earth}}}$ 
\\
Name & [J m$^{-2}$]& [J m$^{-2}$] & [J m$^{-2}$] & [J m$^{-2}$] & & [J m$^{-2}$] &  
\\
& [8]  & [9] & [10] & [11] & [12] & [13] & [14] \\
\hline
GJ 699 b & 1.65E+06 & 8.68E+07 & 7.79E+08 & 2.79 E+04 & 0.16 & 1.66E+06 & 0.00 \\
Kepler-283 c & 4.25E+08 & 1.14E+10 & 2.83E+10 & 1.19E+06 & 6.82 & 4.25E+08 & 0.13 \\ 
Kepler-1634 b &  1.79E+09 & 1.14E+10 & 1.33E+10 & 6.90E+05 & 3.96 & 1.79E+0.9 & 0.53 \\ 
Proxima Cen b & 2.44E+07 & 8.78E+08 & 2.74E+10 & 1.33E+07 & 76.21 & 2.44E+07 & 0.01 \\ 
Ross-128 b & 9.25E+07 & 5.01E+09 & 5.81E+10 & 1.50E+06 & 8.61 & 9.25E+07 & 0.03 \\ 
TRAPPIST-1 b & 2.13E+08 & 8.97E+09 & 1.72E+11 & 7.29E+07 & 418.36 & 2.13E+08 & 0.06\\ 
TRAPPIST-1 c & 1.14E+08 & 4.79E+09 & 9.19E+10 & 3.89E+07 & 223.21 & 1.14E+08 & 0.03   \\ 
TRAPPIST-1 d & 5.73E+07 & 2.41E+09 & 4.63E+10 & 1.96E+07 & 112.34 & 5.73E+07 & 0.02  \\ 
TRAPPIST-1 e & 3.32E+07 & 1.40E+09 & 2.68E+10 & 1.13E+07 & 65.07 & 3.32E+07 & 0.01 \\ 
TRAPPIST-1 f & 1.91E+07 & 8.05E+08 & 1.54E+10 & 6.54E+06 & 37.52 & 1.91E+07 & 0.01  \\ 
TRAPPIST-1 g & 1.30E+07 & 5.44E+08 & 1.05E+10 & 4.42E+06 & 25.39 & 1.30E+07 & 0.00  \\ 
TRAPPIST-1 h & 6.64E+06 & 2.79E+08 & 5.36E+09 & 2.27E+06 & 13.01 & 6.64E+06 & 0.00  \\ 
Sol d (Earth) & 3.37E+09 & 1.93E+10 & 2.04E+10 & 1.74E+05 & 1.00 & 3.37E+09 & 1.00 \\ 
Sol e (Mars) & 1.45E+09 & 8.32E+09 & 8.81E+09 & 7.51E+04 & 0.43 & 1.45E+09 & 0.43  \\ 
\hline
\multicolumn{4}{l}{[1] UV Energy by Annual Maximum Flare at TOA }&
\multicolumn{4}{l}{[2] Ratio to Earth's Annual Maximum Flare}\\
\multicolumn{4}{l}{[3] Ratio to Earth's annual UV flux at TOA }&
\multicolumn{4}{l}{[4] XUV Energy by Annual Maximum Flare at TOA} \\
\multicolumn{4}{l}{[5] Ratio to Earth's Annual Maximum Flare }&
\multicolumn{4}{l}{[6] Ratio to Earth's annual UV flux at TOA}\\
\multicolumn{8}{l}{[7] Annual XUV Energy by Normal Stellar Radiation at TOA }\\
\multicolumn{8}{l}{[8] Annual UV Energy by Normal Stellar Radiation at TOA}\\
\multicolumn{8}{l}{[9] Annual Visible Ray Energy by Normal Stellar Radiation at TOA} \\
\multicolumn{8}{l}{[10] Annual IR Energy by Normal Stellar Radiation at TOA }\\
\multicolumn{8}{l}{[11] Annual Total (flare + Quiescent) XUV Energy at TOA}\\
\multicolumn{8}{l}{[12] Ratio to Earth / Annual Total (flare + Quiescent) XUV Energy at TOA}  \\
\multicolumn{8}{l}{[13] Annual Total (flare + Quiescent) UV Energy at TOA }\\
\multicolumn{8}{l}{[14]Ratio to Earth / Annual Total (flare + Quiescent)　UV Energy at TOA} \\
\multicolumn{8}{l}{TOA - Top of Atmosphere ($\approx$ 0 g/cm$^2$)}
\end{tabular}
\end{table}

\begin{figure}
\begin{center}
\includegraphics[width=1\textwidth]{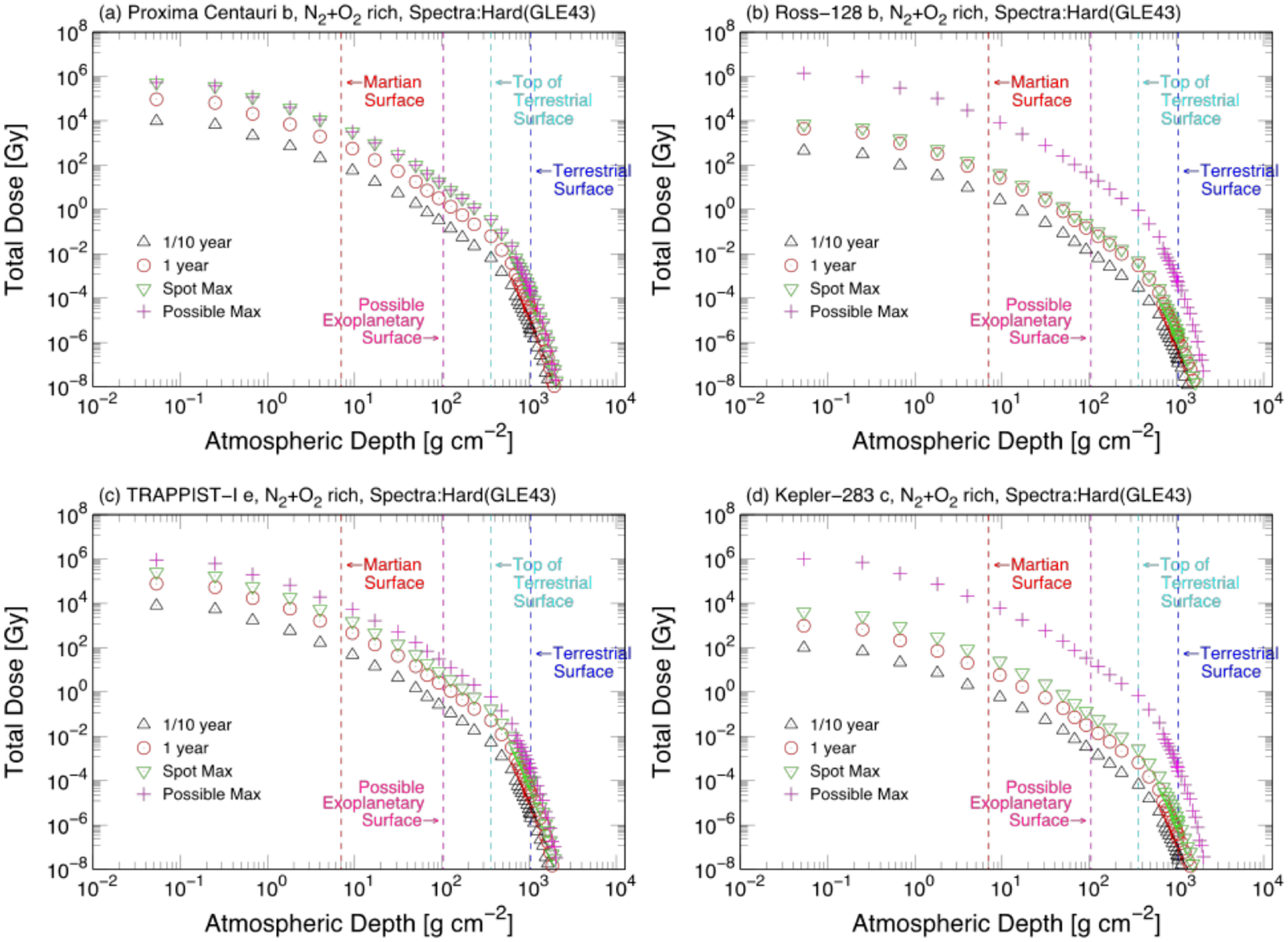}
\caption{Vertical profile of radiation dose (Gy) on Proxima Centauri b, Ross-128 b, TRAPPIST-I and Kepler-283 c for possible flares on several different scales, caused by hard proton spectrum (imitating GLE 43) , penetrating N$_2$ + O$_2$ rich (terrestrial type) atmosphere Earth with flares in every 1/10 year (36 days, black triangle), one year (red circle), spot maximum (green triangle), possible max (pink cross). Martian Surface Atmospheric Pressure, equivalent to 9 g/cm$^2$; Terrestrial Minimum Atmospheric Pressure, observed at the summit of the Himalayas equivalent to 365 g/cm$^2$; (Earth's) Ground Level Atmospheric Pressure, equivalent to 1037 g/cm$^2$;  Possible Exoplanetary Surface, 1/10 of terrestrial surface equivalent to 103.7 g/cm$^2$.}
\label{fig:pf-4planet-Gy}
\end{center}
\end{figure}

\begin{figure}
\begin{center}
\includegraphics[width=1\textwidth]{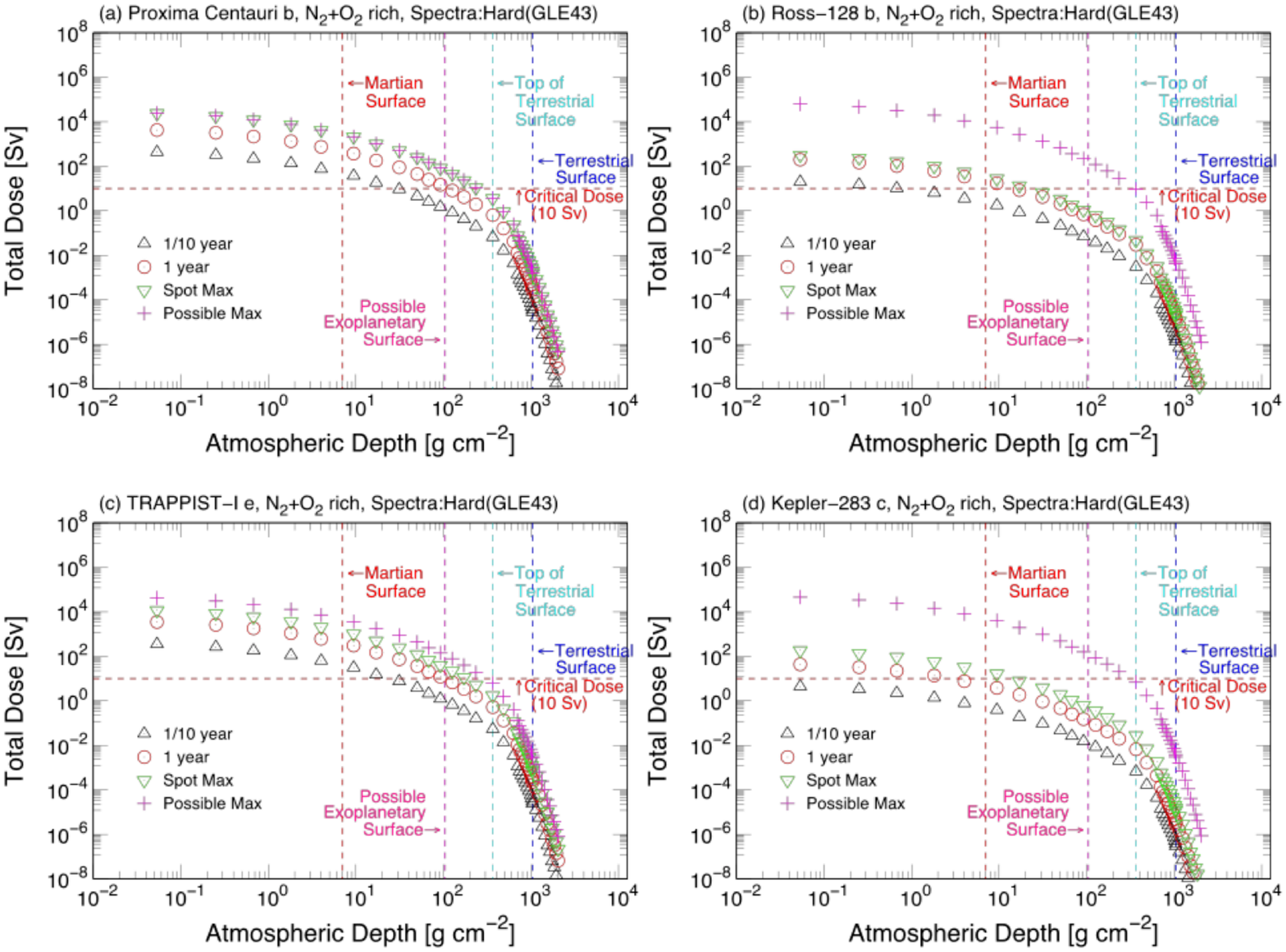}
\caption{Vertical profile of radiation dose (Sv) on Proxima Centauri b, Ross-128 b TRAPPIST-I and Kepler-283 c for possible flares on several different scales, caused by hard proton spectrum (imitating GLE 43) (a)(c) and soft spectrum (Townsend Carrington) (b)(d) penetrating  N$_2$ + O$_2$ rich (terrestrial type) atmosphere Earth with flares every 1/10 year (36 days, black triangle), one year (red circle), spot maximum (green triangle), possible max (rose cross). Martian Surface Atmospheric Pressure, equivalent to 9 g/cm$^2$; Terrestrial Minimum Atmospheric Pressure, observed at the summit of the Himalayas equivalent to 365 g/cm$^2$; (Earth's) Ground Level Atmospheric Pressure, equivalent to 1037 g/cm$^2$;  Possible Exoplanetary Surface, 1/10 of terrestrial surface equivalent to 103.7 g/cm$^2$. }
\label{fig:pf-4planet-Sv}
\end{center}
\end{figure}

Figures \ref{fig:pf-4planet-Gy} and \ref{fig:pf-4planet-Sv} illustrate the vertical radiation dose in Gray (\ref{fig:pf-4planet-Gy}) and in Sievert(\ref{fig:pf-4planet-Sv}) for major documented planetary systems, including Proxima Centauri b, Ross-128 b, TRAPPIST-1 e and Kepler-283 c (the only habitable planet in the Kepler field with observed flares)  for possible flares on several different scales, caused by hard proton spectrum (imitating GLE 43) penetrating N$_2$+O$_2$  rich (terrestrial type) atmosphere Earth with flares every 1/10 year ( 36 days), one year (Annual), Spot Maximum, and Possible Maximum flare.  In these scenarios, the Spot Maximum flare is the maximum possible flare to be observed within decades in the target stellar system. On these planets, the SPE does not reach critical levels with sufficient atmospheric depth (equal to that of the Earth) even for the Possible Maximum Flare scenario, having 1.42 $\times 10^{-4}$ Gy (1.6 $\times 10^{-3}$ Sv) for Proxima Centauri b and  3.70 $\times 10^{-4}$ Gy (4.14 $\times 10^{-3}$ Sv) for Ross-128 b. Proxima Centauri b shows a smaller difference between the Possible Maximum Flare dose and Spot Maximum Flare dose, whereas a larger difference can be found on Ross-128 b, showing that Ross-128b is relatively calm in the range of the same temperature class. 

However, when considering the Possible Maximum Flare, calculated assuming that the whole star is covered by the maximum percentage of starspot (20\%, observed from Proxima Centauri's light-curve survey), the radiation dose at the terrestrial lowest atmospheric thickness measured at the summit of Everest (at AD 365 g/cm$^2$ set in this study) applied to Proxima Centauri b, Ross-128 b, TRAPPIST-1 e, and Kepler-283 c, the estimated dose on these planets reach a fatal dose, of  0.36 Gy (3.64 Sv), 0.93 Gy (9.45 Sv), 3.03 Gy(30.8 Sv), and 0.68 Gy(6.89 Sv), respectively.

\begin{figure}
\begin{center}
\includegraphics[width=1\textwidth]{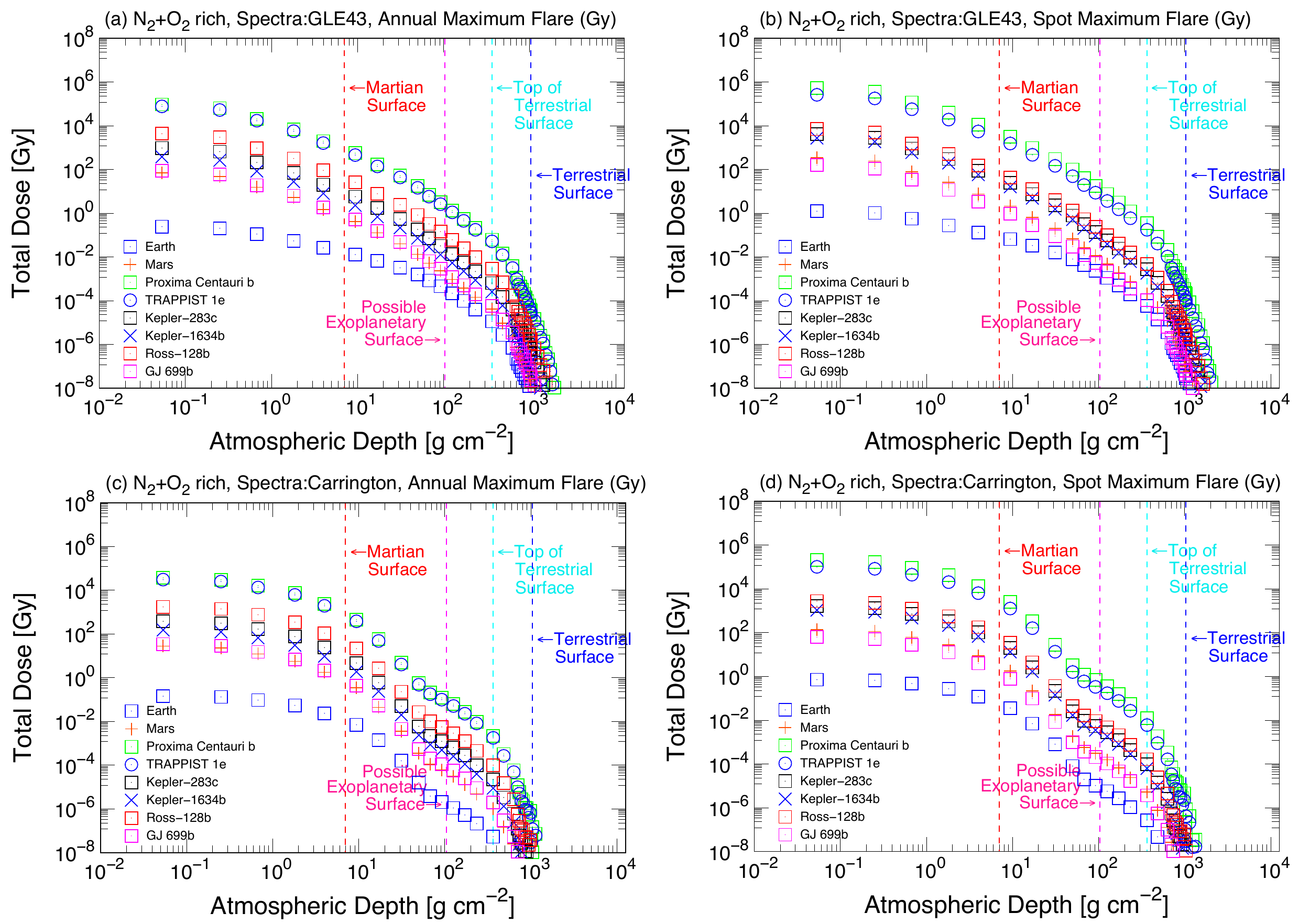}
\caption{Vertical profile of radiation dose (Gy), caused by proton spectrum imitating GLE 43 (a)(b) and Carrington Flare (c)(d) penetrating N$_2$ + O$_2$ rich (terrestrial type) atmosphere on Proxima Centauri b (green square), TRAPPIST-1 e (blue circle), Kepler-283 c (brown square), Kepler-1634 b (blue cross), Ross-128 b (red square) and GJ-699 b (pink square) in comparison with the Earth (blue square) and Mars (red plus) in logarithmic scale under Annual Maximum flare energy  (a) (c) under Spot Maximum flare energy (b)(d), in Gray (Gy). Martian Surface Atmospheric Pressure, equivalent to 9 g/cm$^2$; Terrestrial Minimum Atmospheric Pressure, observed at the summit of the Himalayas equivalent to 365 g/cm$^2$; (Earth's) Ground Level Atmospheric Pressure, equivalent to 1037 g/cm$^2$;  Possible Exoplanetary Surface, 1/10 of terrestrial surface equivalent to 103.7 g/cm$^2$.}
\label{fig:NOGy}
\end{center}
\end{figure}

\begin{figure}
\begin{center}
\includegraphics[width=1\textwidth]{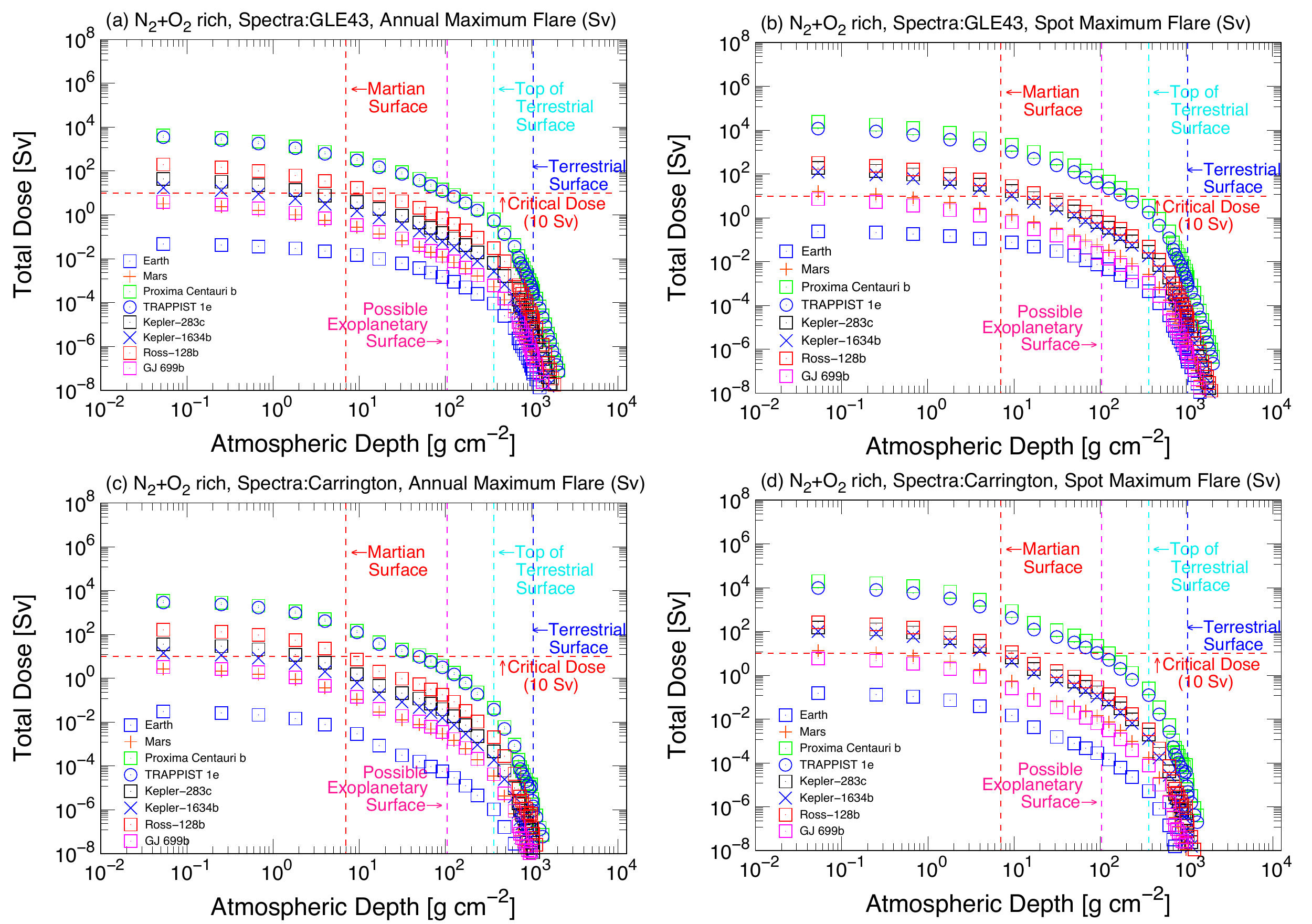}
\caption{Vertical profile of radiation dose (Sv), caused by proton spectrum imitating GLE 43 (a)(b) and Carrington Flare (c)(d) penetrating N$_2$ + O$_2$ rich (terrestrial type) atmosphere on Proxima Centauri b (green square), TRAPPIST-1 e (blue circle), Kepler-283 c (brown square), Kepler-1634 b (blue cross), Ross-128 b (red square) and GJ-699 b (pink square)  in comparison with the Earth (blue square) and Mars (red plus) in logarithmic scale under Annual Maximum flare energy  (a) (c) under Spot Maximum flare energy (b)(d), in Sievert (Sv). Martian Surface Atmospheric Pressure, equivalent to 9 g/cm$^2$; Terrestrial Minimum Atmospheric Pressure, observed at the summit of the Himalayas equivalent to 365 g/cm$^2$; (Earth's) Ground Level Atmospheric Pressure, equivalent to 1037 g/cm$^2$;  Possible Exoplanetary Surface, 1/10 of terrestrial surface equivalent to 103.7 g/cm$^2$.}
\label{fig:NOSv}
\end{center}
\end{figure}

We calculated the vertical profile of the radiation dose, caused by the proton spectrum similar to the one reconstructed for the GLE 43 and Carrington-class events for each planet (GJ-699b, Kepler-283c, Kepler-1634 b, Proxima Centauri b, Proxima Centauri b, Ross 128 b, TRAPPIST-1 e) with terrestrial-type atmospheric compositions under Annual Maximum Flare and Spot Maximum Flare events, by comparing that of Earth and Mars (see Figures \ref{fig:NOGy} and \ref{fig:NOSv}).  

\begin{figure}
\begin{center}
\includegraphics[width=1\textwidth]{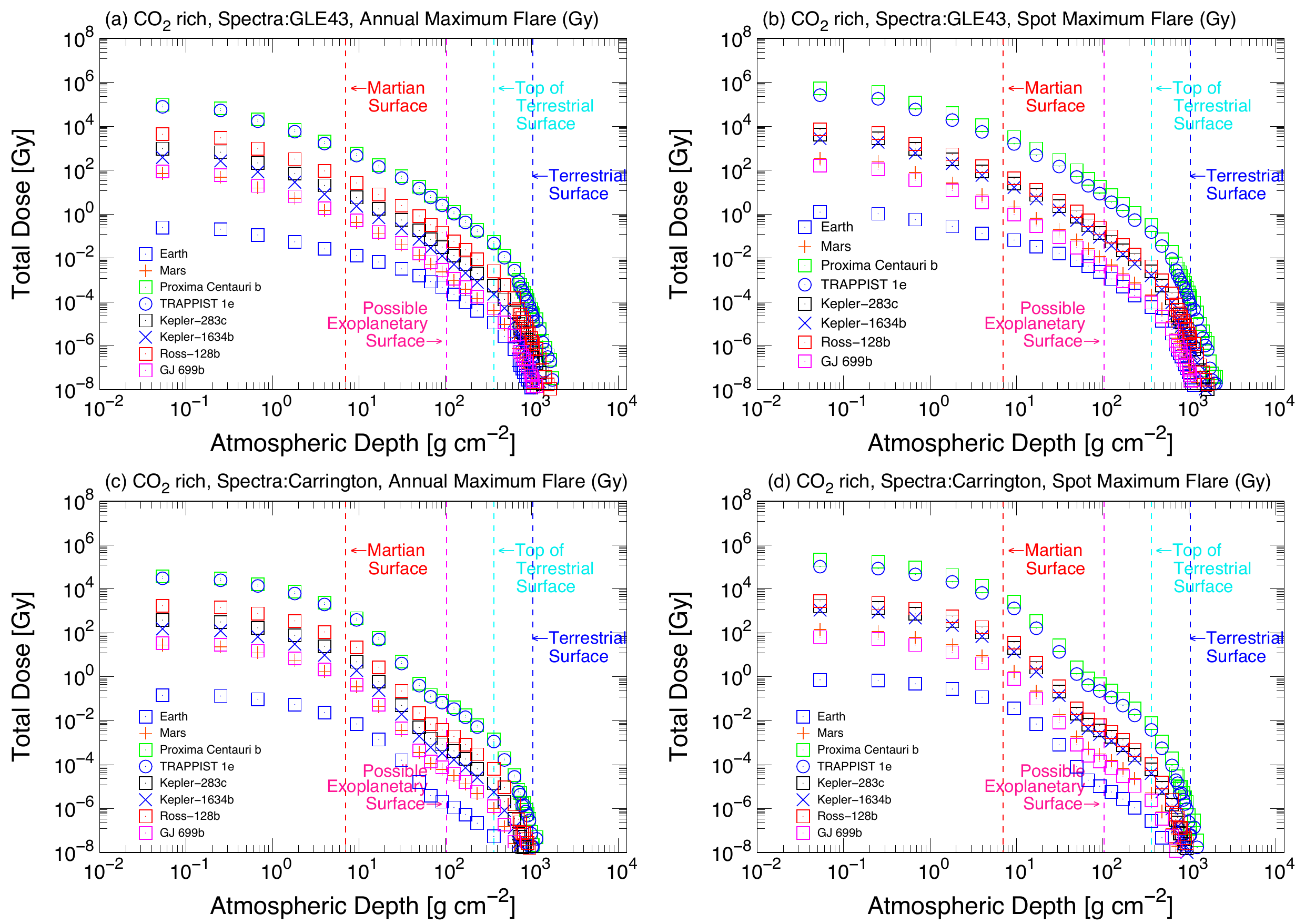}
\caption{Vertical profile of radiation dose (Gy), caused by proton spectrum imitating GLE 43 (a)(b) and Carrington Flare (c)(d) penetrating CO$_2$ rich (terrestrial type) atmosphere on Proxima Centauri b (green square), TRAPPIST-1 e (blue circle), Kepler-283 c (brown square), Kepler-1634 b (blue cross), Ross-128 b (red square) and GJ-699 b (pink square)  in comparison with the Earth (blue square) and Mars (red plus) in logarithmic scale under Annual Maximum flare energy  (a) (c) under Spot Maximum flare energy  (b)(d), in Gray (Gy).Martian Surface Atmospheric Pressure, equivalent to 9 g/cm$^2$; Terrestrial Minimum Atmospheric Pressure, observed at the summit of the Himalayas equivalent to 365 g/cm$^2$; (Earth's) Ground Level Atmospheric Pressure, equivalent to 1037 g/cm$^2$;  Possible Exoplanetary Surface, 1/10 of terrestrial surface equivalent to 103.7 g/cm$^2$.}
\label{fig:CO2Gy}
\end{center}
\end{figure}

\begin{figure}
\begin{center}
\includegraphics[width=1\textwidth]{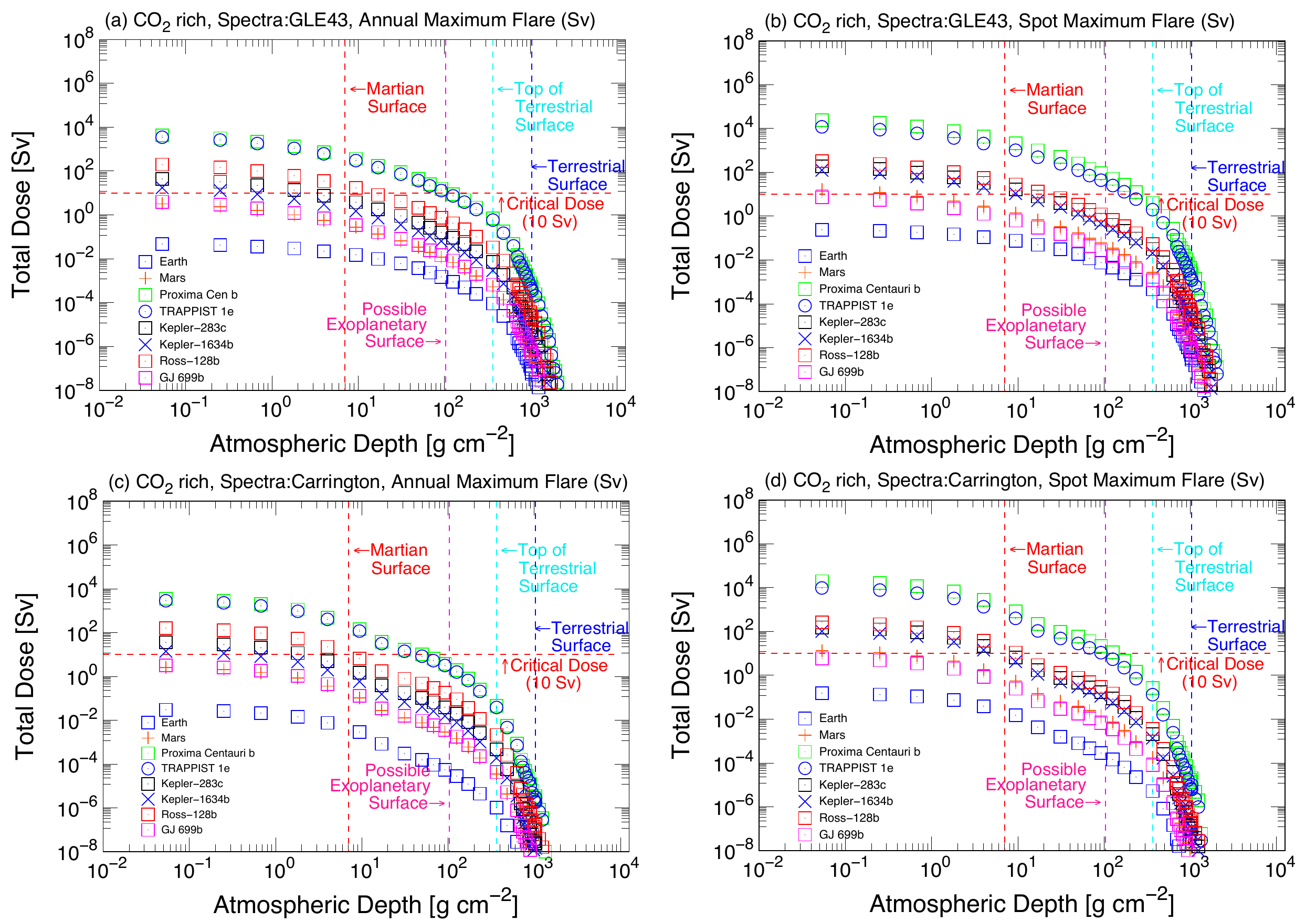}
\caption{Vertical profile of radiation dose (Sv), caused by proton spectrum imitating GLE 43 (a)(b) and Carrington Flare (c)(d) penetrating CO$_2$ rich (terrestrial type) atmosphere on Proxima Centauri b (green square), TRAPPIST-1 e (blue circle), Kepler-283 c (brown square), Kepler-1634 b (blue cross), Ross-128 b (red square) and GJ-699 b (pink square)  in comparison with the Earth (blue square) and Mars (red plus) in logarithmic scale under Annual Maximum flare energy  (a) (c) under Spot Maximum flare energy  (b)(d), in Sievert (Sv). Martian Surface Atmospheric Pressure, equivalent to 9 g/cm$^2$; Terrestrial Minimum Atmospheric Pressure, observed at the summit of the Himalayas equivalent to 365 g/cm$^2$; (Earth's) Ground Level Atmospheric Pressure, equivalent to 1037 g/cm$^2$;  Possible Exoplanetary Surface, 1/10 of terrestrial surface equivalent to 103.7 g/cm$^2$.}
\label{fig:CO2Sv}
\end{center}
\end{figure}

\begin{figure}
\begin{center}
\includegraphics[width=1\textwidth]{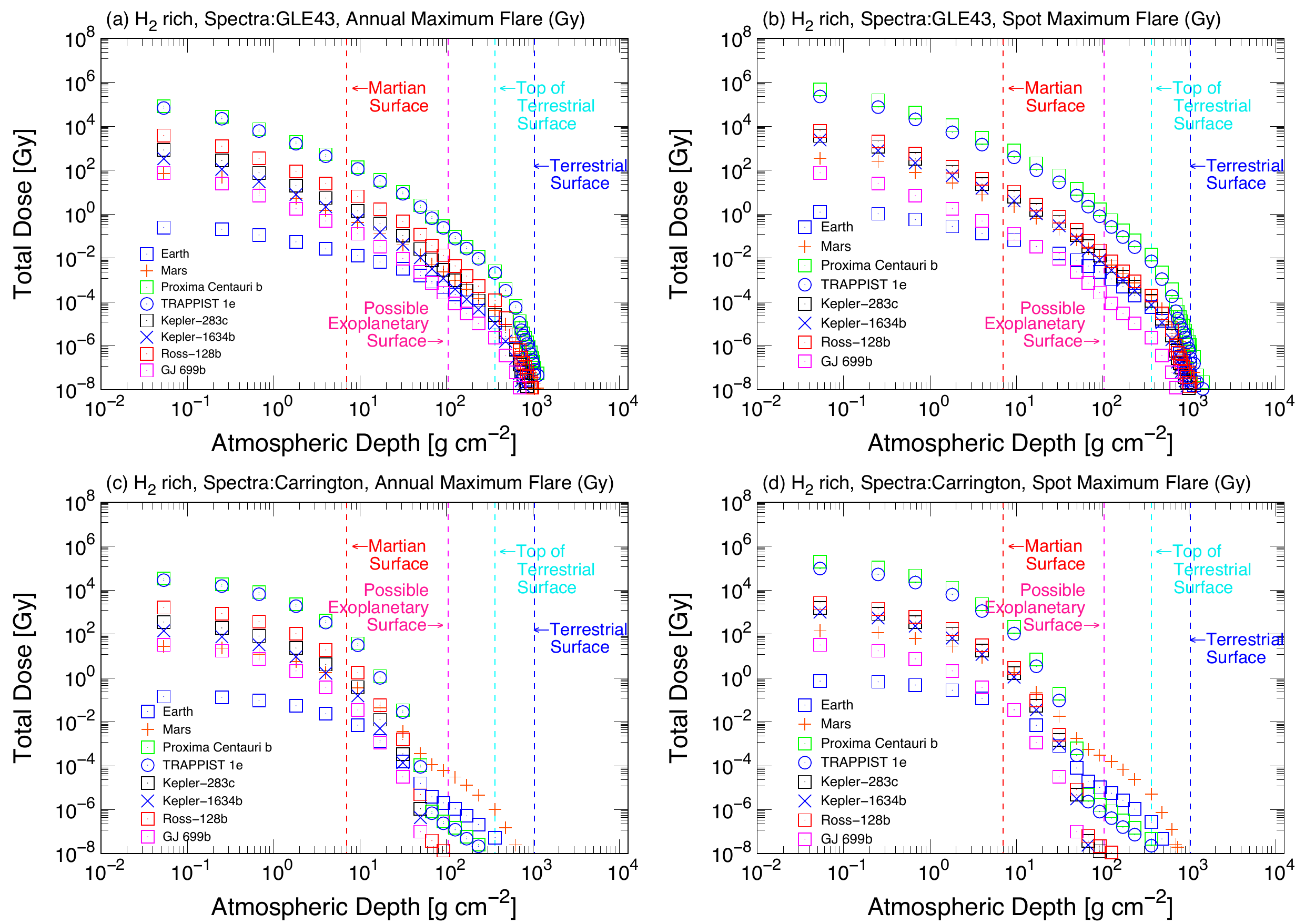}
\caption{Vertical profile of radiation dose (Sv), caused by proton spectrum imitating GLE 43 (a)(b) and Carrington Flare (c)(d) penetrating H$_2$ rich (terrestrial type) atmosphere on Proxima Centauri b (green square), TRAPPIST-1 e (blue circle), Kepler-283 c (brown square), Kepler-1634 b (blue cross), Ross-128 b (red square) and GJ-699 b (pink square)  in comparison with the Earth (blue square) and Mars (red plus) in logarithmic scale under Annual Maximum flare energy  (a) (c) under Spot Maximum flare energy  (b)(d), in Gray (Gy). Martian Surface Atmospheric Pressure, equivalent to 9 g/cm$^2$; Terrestrial Minimum Atmospheric Pressure, observed at the summit of the Himalayas equivalent to 365 g/cm$^2$; (Earth's) Ground Level Atmospheric Pressure, equivalent to 1037 g/cm$^2$;  Possible Exoplanetary Surface, 1/10 of terrestrial surface equivalent to 103.7 g/cm$^2$.}
\label{fig:H2Gy}
\end{center}
\end{figure}

\begin{figure}
\begin{center}
\includegraphics[width=1\textwidth]{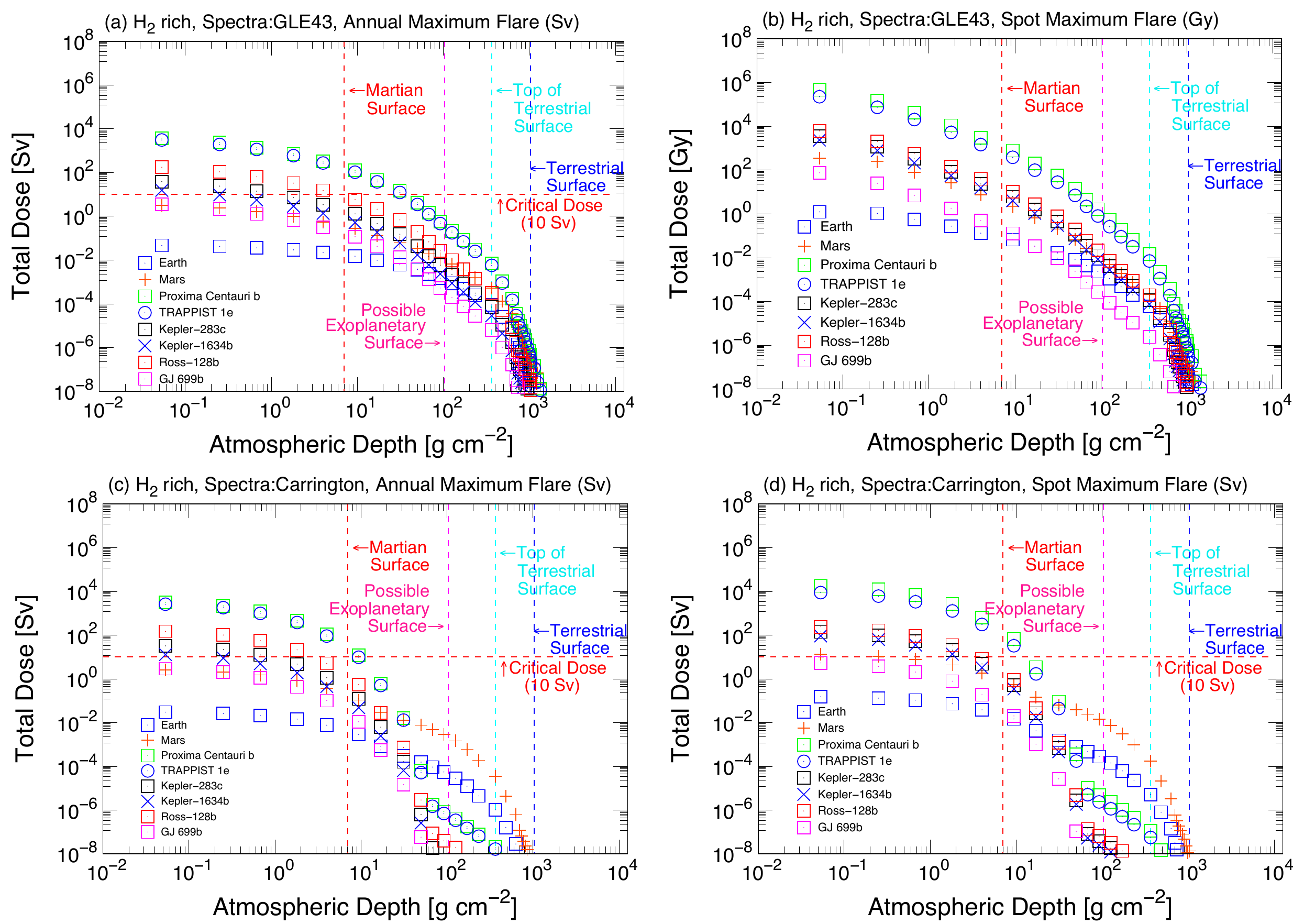}
\caption{Vertical profile of radiation dose (Sv), caused by proton spectrum imitating GLE 43 (a)(b) and Carrington Flare (c)(d) penetrating H$_2$ rich (terrestrial type) atmosphere on Proxima Centauri b (green square), TRAPPIST-1 e (blue circle), Kepler-283 c (brown square), Kepler-1634 b (blue cross), Ross-128 b (red square) and GJ-699 b (pink square)  in comparison with the Earth (blue square) and Mars (red plus) in logarithmic scale under Annual Maximum flare energy  (a) (c) under Spot Maximum flare energy  (b)(d), in Sievert (Sv). Martian Surface Atmospheric Pressure, equivalent to 9 g/cm$^2$; Terrestrial Minimum Atmospheric Pressure, observed at the summit of the Himalayas equivalent to 365 g/cm$^2$; (Earth's) Ground Level Atmospheric Pressure, equivalent to 1037 g/cm$^2$;  Possible Exoplanetary Surface, 1/10 of terrestrial surface equivalent to 103.7 g/cm$^2$.}
\label{fig:H2Sv}
\end{center}
\end{figure}

\begin{figure}
\begin{center}
\includegraphics[width=1\textwidth]{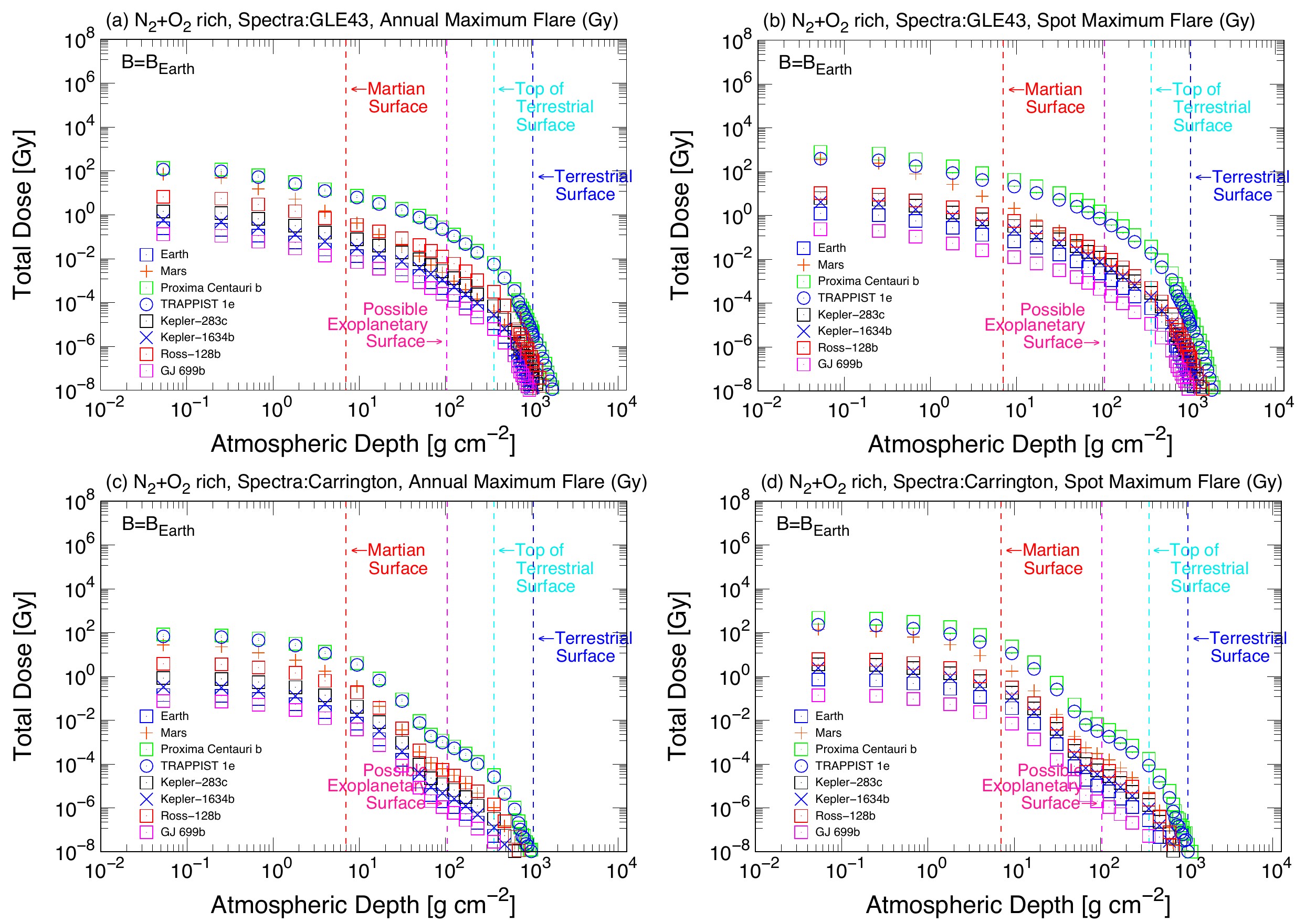}
\caption{Vertical profile of radiation dose (Sv), caused by proton spectrum imitating GLE 43 (a)(b) and Carrington Flare (c)(d) penetrating H$_2$ rich (terrestrial type) atmosphere on Proxima Centauri b (green square), TRAPPIST-1 e (blue circle), Kepler-283 c (brown square), Kepler-1634 b (blue cross), Ross-128 b (red square) and GJ-699 b (pink square)  in comparison with the Earth (blue square) and Mars (red plus) in logarithmic scale under Annual Maximum flare energy  (a) (c) under Spot Maximum flare energy  (b)(d), in Gray (Gy). Martian Surface Atmospheric Pressure, equivalent to 9 g/cm$^2$; Terrestrial Minimum Atmospheric Pressure, observed at the summit of the Himalayas equivalent to 365 g/cm$^2$; (Earth's) Ground Level Atmospheric Pressure, equivalent to 1037 g/cm$^2$;  Possible Exoplanetary Surface, 1/10 of terrestrial surface equivalent to 103.7 g/cm$^2$.}
\label{fig:NOGy1B}
\end{center}
\end{figure}

\begin{figure}
\begin{center}
\includegraphics[width=1\textwidth]{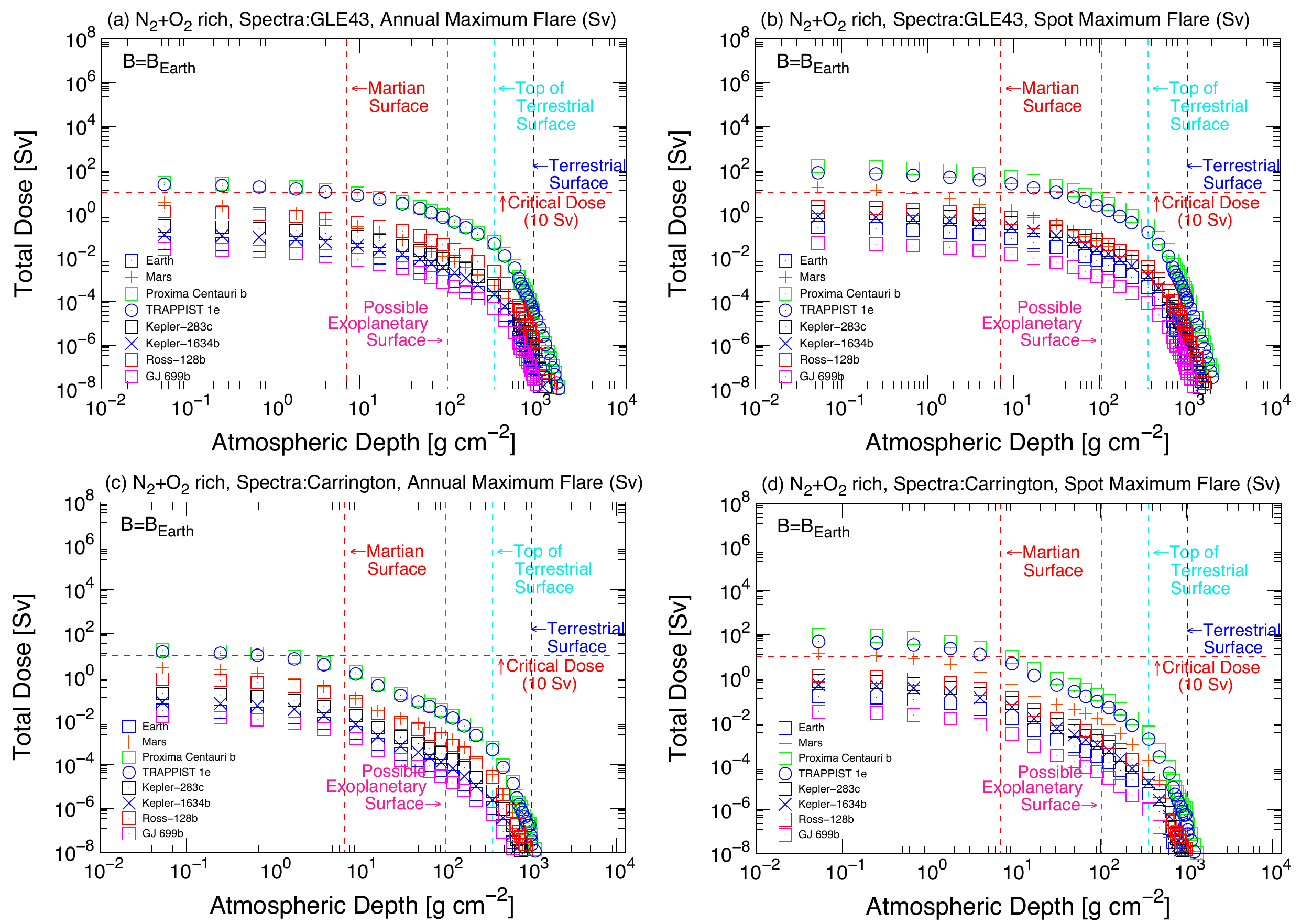}
\caption{Vertical profile of radiation dose (Sv), caused by proton spectrum imitating GLE 43 (a)(b) and Carrington Flare (c)(d) penetrating H$_2$ rich (terrestrial type) atmosphere on Proxima Centauri b (green square), TRAPPIST-1 e (blue circle), Kepler-283 c (brown square), Kepler-1634 b (blue cross), Ross-128 b (red square) and GJ-699 b (pink square)  in comparison with the Earth (blue square) and Mars (red plus) in logarithmic scale under Annual Maximum flare energy  (a) (c) under Spot Maximum flare energy  (b)(d), in Sievert (Sv).Martian Surface Atmospheric Pressure, equivalent to 9 g/cm$^2$; Terrestrial Minimum Atmospheric Pressure, observed at the summit of the Himalayas equivalent to 365 g/cm$^2$; (Earth's) Ground Level Atmospheric Pressure, equivalent to 1037 g/cm$^2$;  Possible Exoplanetary Surface, 1/10 of terrestrial surface equivalent to 103.7 g/cm$^2$.}
\label{fig:NOSv1B}
\end{center}
\end{figure}

For the evaluation at different atmospheric depths we employed the following four typical atmospheric depth reference layers: Top of Atmosphere  (TOA) equivalent to $\approx$ 0 g/cm$^2$, Martian Surface Atmospheric Pressure (MS) equivalent to 9 g/cm$^2$, Terrestrial Minimum Atmospheric Pressure, observed at the summit of the Himalayas equivalent to 365 g/cm$^2$ in this study, and (Earth's) Ground Level Atmospheric Pressure, equivalent to 1037 g/cm$^2$.  Possible Exoplanetary Surface was estimated as 1/10 of the terrestrial surface equivalent to 103.7 g/cm$^2$. Note that the value is not identical to the real observation data but at the nearest value employed in the Monte-Carlo numerical simulation. 

According to these calculations, we can specify the critical dose for each planet, which is presumed in this study to be 10 Sv per annual event (see Radiation Dose subsection). Using this threshold, we may determine the minimum requirement of the atmospheric depth for terrestrial-type lifeform evolution.  According to our analysis, the critical atmospheric depths required to secure terrestrial-type lifeform evolution on the surface of each modeled exoplanet exposed by annual severe flare events are: 2.77 g/cm$^2$ for GJ 699 b (Barnard's Star b) (0.267 \% of terrestrial atmospheric depth) , 3.27 $\times 10^2$ g/cm$^2$ for Proxima Cen b (31.6 \% of terrestrial atmospheric depth) , 7.59 $\times 10$ g/cm$^2$ (7.31 \% ) for Ross 128 b, and 3.06 $\times 10^2$ g/cm$^2$ for TRAPPIST-1 e (29.5 \%)(Figure \ref{fig:NOGy} and \ref{fig:NOSv}).
We note that without sufficient atmospheric depth, the surface primitive lifeforms on those planets suffer from severe radiation doses, even for relatively small-scale flares. 

We also performed calculations of the radiation doses for CO$_{2}$ rich and H$_{2}$ rich atmospheres for each planet (see Figures \ref{fig:CO2Gy}, \ref{fig:CO2Sv}, \ref{fig:H2Gy} and \ref{fig:H2Sv}). The difference between each atmospheric composition does not become significant especially when compared with N$_2$+O$_2$ and CO$_{2}$ rich type. However, it is evident that H$_{2} $ rich atmosphere dissipates higher energetic particles more significantly. 

\begin{figure}
\begin{center}
\includegraphics[width=1\textwidth]{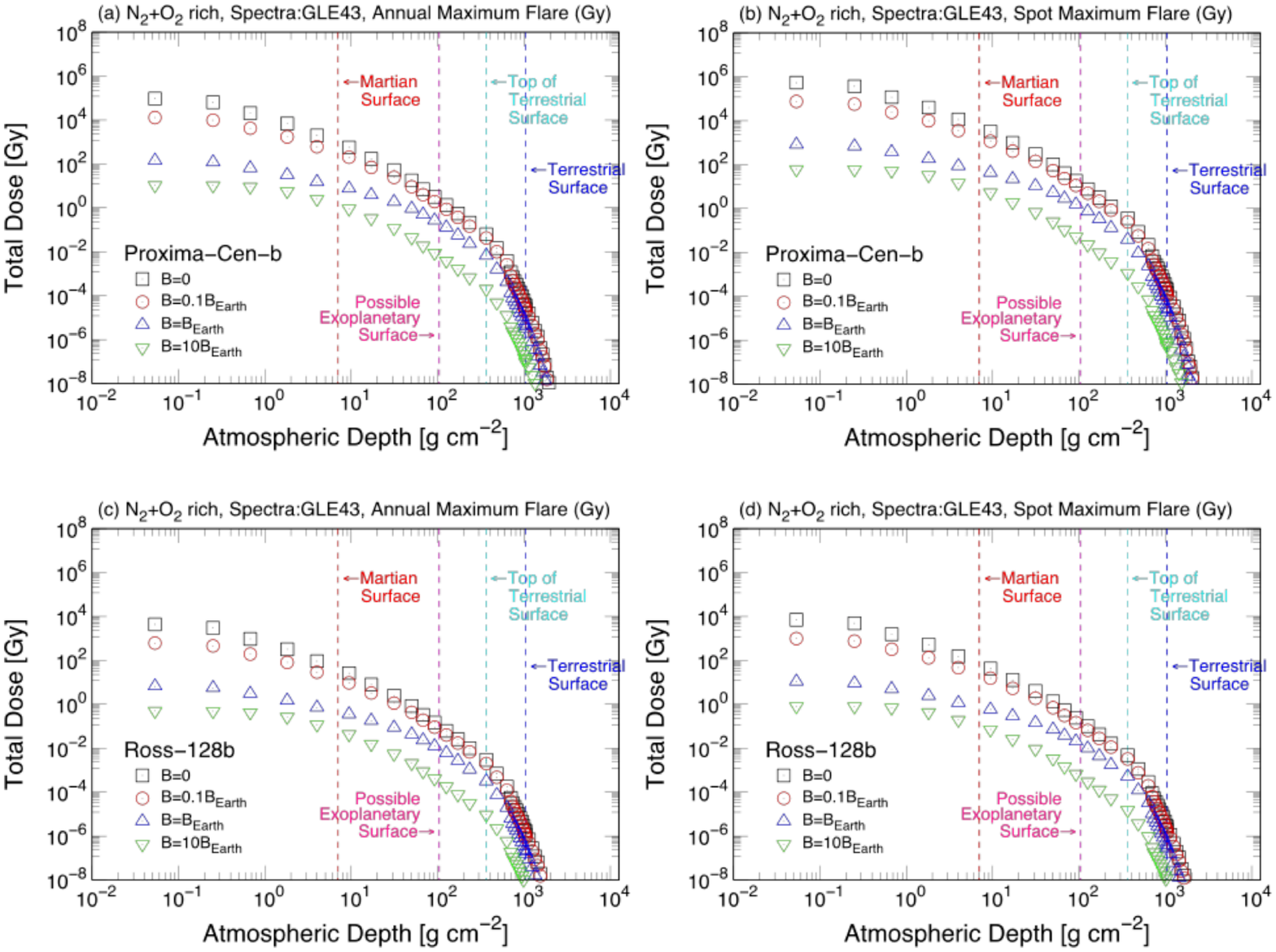}
\caption{Vertical distribution of each radiation dose (Gy) under different magnetic shower caused by SPE Air Shower penetrating N$_2$ + O$_2$ rich (terrestrial type) on Proxima Centauri b (a)(b) and on Ross 128 b (c)(d) in logarithmic scale under Annual Maximum flare energy  (a) (c) under Spot Maximum flare energy calculated by Shibata et al. 2013(b)(d). Martian Surface Atmospheric Pressure, equivalent to 9 g/cm$^2$; Terrestrial Minimum Atmospheric Pressure, observed at the summit of the Himalayas equivalent to 365 g/cm$^2$; (Earth's) Ground Level Atmospheric Pressure, equivalent to 1037 g/cm$^2$;  Possible Exoplanetary Surface, 1/10 of terrestrial surface equivalent to 103.7 g/cm$^2$. }
\label{fig:difmag-PR}
\end{center}
\end{figure}

The presence of a geomagnetic dipole shield around a planet and its relative strength will influence its efficacy for reducing the irradiation effect of any solar flares on surface lifeforms. First, we recalculated the surface dose (Gy and Sv) assuming that all exoplanets (GJ-699 b, Kepler-283 c, Kepler-1634 b, Proxima Centauri b, Ross-128 b, TRAPPIST-1 e in comparison with Earth \& Mars) have the same amount of magnetic shield as the Earth (B = B$_{Earth}$) (see Figure \ref{fig:NOGy1B} and \ref{fig:NOSv1B}). Then we evaluated the scenarios with the planetary dipole magnetic field (uniform over the whole planet surface) of (i) 0 (no magnetosphere), (ii) $0.1 \times B_{\mathrm{Earth}}$, (iii) $1 \times  B_{\mathrm{Earth}}$ (Earth level),  and (iv) $10 \times B_{\mathrm{Earth}}$ for four documented exoplanets (Proxima Centauri b, Ross-128b, and TRAPPIST-1 e and Kepler-283 c) (see Figure \ref{fig:difmag-PR}). 

\begin{figure}
\begin{center}
\includegraphics[width=1\textwidth]{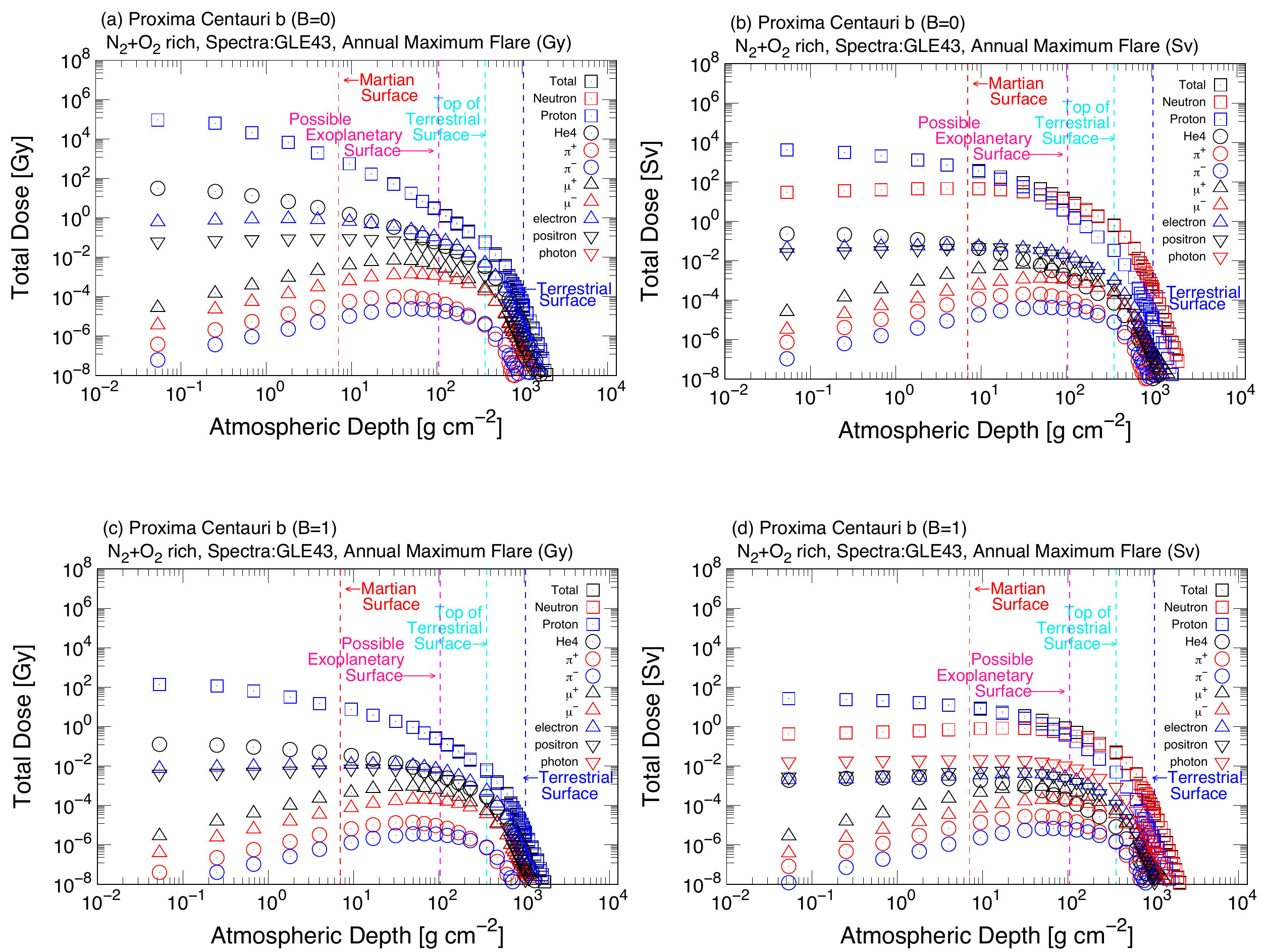}
\caption{Vertical profile of radiation dose in Gray and Sievert, caused by SPE Air Shower penetrating N$_2$ + O$_2$ rich (terrestrial type) atmosphere on Proxima Centauri b with B=0 (a) (b) and with B=Bearth(c) (d) under annual maximum flare energy in Gray (a)(c) and Sievert (b)(d).Martian Surface Atmospheric Pressure, equivalent to 9 g/cm$^2$; Terrestrial Minimum Atmospheric Pressure, observed at the summit of the Himalayas equivalent to 365 g/cm$^2$; (Earth's) Ground Level Atmospheric Pressure, equivalent to 1037 g/cm$^2$;  Possible Exoplanetary Surface, 1/10 of terrestrial surface equivalent to 103.7 g/cm$^2$.}
\label{fig:difmag-shower-P}
\end{center}
\end{figure}

\begin{figure}
\begin{center}
\includegraphics[width=1\textwidth]{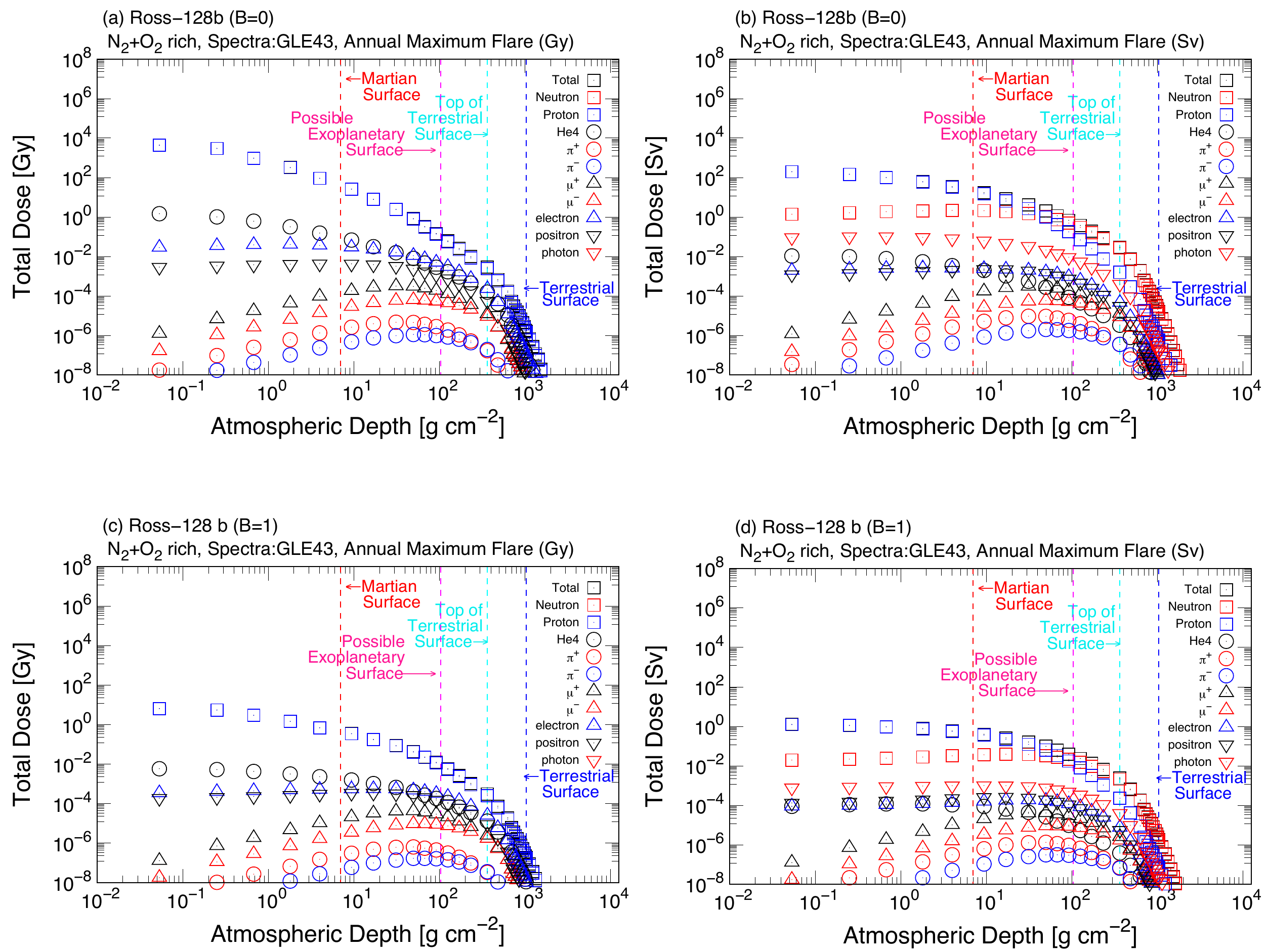}
\caption{Vertical profile of radiation dose in Gray and Sievert, caused by SPE Air Shower penetratingN$_2$ + O$_2$ rich (terrestrial type) atmosphere on Ross-128 b with B=0 (a) (b) and with B=Bearth(c) (d) under annual maximum flare energy  in Gray (a)(c) and Sievert (b)(d). Martian Surface Atmospheric Pressure, equivalent to 9 g/cm$^2$; Terrestrial Minimum Atmospheric Pressure, observed at the summit of the Himalayas equivalent to 365 g/cm$^2$; (Earth's) Ground Level Atmospheric Pressure, equivalent to 1037 g/cm$^2$;  Possible Exoplanetary Surface, 1/10 of terrestrial surface equivalent to 103.7 g/cm$^2$.}
\label{fig:difmag-shower-R}
\end{center}
\end{figure}

\subsection{Atmospheric Escape induced by XUV Flux and Associated Possible Higher Radiation Dose}
XUV radiation caused by stellar superflares does not significantly increase the annual planetary surface UV flux on the exoplanets modeled unless high stellar magnetically driven events (flares and CMEs) severely damage their atmospheric thickness and impact their chemistry. However, since stellar quiescent XUV radiation induces atmospheric escape from a terrestrial-type planet, a more critical situation may be predicted; that the atmospheric depth of those exoplanets easily reaches at least 1/10 of terrestrial atmospheric thickness. The atmospheric escape rate of $O^{+}/N^{+}$ ions from Proxima Centauri b, TRAPPIST-1 e, Ross-128b and Kepler-283c are 76.2, 53.5,  7.92, and 6.82 times stronger than that of the Earth due to higher stellar XUV fluxes incident on the planetary atmospheres caused by closer proximity to their respective host stars according to the equation proposed by \citet{Airapetian2017a}. 

In such a scenario, the radiation dose at Proxima Centauri b and TRAPPIST-1 e reaches nearly fatal levels for complex lifeforms even through the modest annual flares, reaching 1.32 Gy (8.09 Sv), and 1.09 Gy (6.68 Sv), respectively (see Figure \ref{fig:pf-4planet-Sv}).

\section{Discussion and Future Work}

\subsection{Estimated dose by Stellar Proton Event in Exoplanets}
The Stellar Proton Event impact onto different exoplanets has been evaluated assuming three types of major atmosphere (N$_2$ + O$_2$, CO$_2$, H$_2$). In general, H$_2$ rich atmosphere, which may be present either younger Earth-sized exoplanets or super-Earths with relatively larger masses,  has a maximum absorption ratio compared with the other two atmospheres. This might be because of the lower molecular weight of Hydrogen, which enables larger molecular numbers under the same atmospheric pressure. Under the other two major types of atmosphere (N$_2$ + O$_2$, CO$_2$), there are still significant reduction effects when atmospheric pressure is sufficient as that of the Earth. For TRAPPIST-1 e and Proxima Centauri b, even for Annual Maximum Flare events, the dose becomes relatively large but not at the level which may affect complex lifeforms. However, with reduced atmospheric pressure to the level at the Top of Terrestrial Surface (352 g/cm$^2$), observed at the summit of the Himalayas, the dose becomes significantly high. The dose with higher atmospheric depth than Martian Surface, especially in the unit in Sieverts, has been evaluated higher than in previous papers calculated by different approaches with hypothetical stellar flare magnitude\citep{Atri2017}. This may be induced by the difference in definitions of the effective doses, as well as precise numerical evaluation using PHITS \citep{Sato2018b}. This relative higher dose in Sievert compared with the unit in Gray, calculated at the higher atmospheric depth than Martian Surface, is mainly induced by the neutron particle, generated as secondary cosmic ray when SEP reaches to the atmosphere, as shown in Figures \ref{fig:difmag-shower-P} and \ref{fig:difmag-shower-R},

Our conclusion is that, with relevant thickness of atmospheric depth for each type of atmosphere, there will be no significant damage to surface lifeforms, except for some critical planets located very close to their stars (TRAPPIST-1 d). 

Projected Stellar Proton Events with harder spectra using GLE43 have a deeper penetration of intensive protons toward the terrestrial atmospheric depth, whereas projected Stellar Proton Events with softer spectra using the Carrington Flare have a sudden reduction of the dose at the mid altitude of the atmosphere (atmospheric depth between $10^{1}$ and $10^{2}$). 

By taking a look at Figure \ref{fig:NOSv1B}, most of the critical dose only applies for Proxima Centauri b and TRAPPIST-1 e when the atmospheric depth was lower than that of Martian Surface when  all exoplanets have the same amount of magnetic shield as Earth (B = B$_{Earth}$), except the scenario with Spot Maximum Flare with proton spectrum imitating GLE 43 (b). 
Each value of the magnetic field may result in a significant dose reduction at the TOA; from (i) 9.37 $\times 10^{4} $ Gy (4.24 $\times 10^{3} $ Sv) with no magnetosphere  to (iii)  1.40  $\times 10^{2} $ Gy (2.7 $\times 10^{1} $ Sv) with Earth level magnetosphere  ($1 \times  B_{\mathrm{Earth}}$ ) almost 1/700 of the dose in Gy (1/160 in Sv) that would be received with no protective magnetic field. Also at the ground level, the dose was reduced from (i) 2.49 $\times 10^{-5} $ Gy (2.79 $\times 10^{-5} $ Sv) with no magnetosphere,  to (iii)  3.23 $\times 10^{-6} $ Gy (3.19 $\times 10^{-5} $ Sv)  with Earth level magnetosphere ($1 \times  B_{\mathrm{Earth}}$ ) on Proxima Cen b, almost 1/10 of the dose that would be received with no protective magnetic field. We may consider that, with the presence of magnetic shields, those listed planets all become habitable at least when the minimum amount of atmospheric depth is present.

\subsection{Impact of XUV radiation}
We also evaluated XUV flux values from stellar flares as follows (see XUV radiation subsection).
Here we assume that the UV energy portion is up to 10\% of the total flare energy,
and estimate XUV dose by annual maximum flare.
As a result, annual XUV dose values due to stellar flares at the TOA
of the target exoplanets
are $10^5 \sim 10^6$ J m$^{-2}$ (see Table \ref{tab:Table2}).
They are all are smaller than 0.001 \% of the terrestrial annual UV dose 
at the TOA ($\sim 4.3\times 10^{9}$ J m$^{-2}$).

In addition to estimating the flare XUV values, 
we also roughly evaluate the quiescent component of XUV fluxes from stellar temperature (spectral class) and spot size 
(See Appendix UV Radiation). 
As a result, the annual total flux of the whole XUV \& UV wavelength range (10-4000 \AA) at the TOA of all our target exoplanets is smaller than the annual total of the Earth. For example, in the most severe case, Kepler-1634b shows
56\% of the Earth values. The impulsive UV doses from annual maximum flares are not significant when compared with the annual dose from steady components.
 In contrast, the XUV (1-1200 \AA) fluxes at the TOA of the target exoplanets have much higher values compared to those at Earth. For example,Proxima Centauri experiences $\sim$76 times larger annual XUV flux at its TOA as compared with 
Earth's value, while TRAPPIST-1e has $\sim$65 times larger flux values.
This is because the XUV contribution in the overall UV emission from cool M-dwarfs are 
larger than that for the Sun \citep{Ribas2017}.

\subsection{Estimation of Atmospheric Escape Rate Due to Photoionization}
Since coronal XUV radiation induces atmospheric escape via photoionization, we can estimate the atmospheric escape rate on Proxima Centauri b, TRAPPIST-1 e, Ross-128b and Kepler-283c using the proposed XUV flux-escape rate scaling by \citet{Airapetian2017a}. In order to estimate these values, we needed to obtain or synthesize the possible XUV flux for those hoststars. In this study, we calculated all XUV fluxes according to the method illustrated in the Appendix. By obtaining the XUV fluxes, we compared them with the atmospheric escape rate from the Earth. The atmospheric escape rates via quiescent XUV emission from Proxima Centauri b, TRAPPIST-1 e, Ross-128b and Kepler-283c are 76.2, 53.5,  7.92, and 6.82 times stronger than that of the Earth due to the higher XUV fluxes incident on close-in exoplanetary atmospheres. If not enough outgassing for these planets are expected, the assumed atmospheric depths especially for Proxima Centauri b and TRAPPIST-1 e may reach $\leqq$ 1/10 of the atmospheric pressure on Earth. In such a scenario, the radiation dose on Proxima Centauri b and TRAPPIST-1 e reach nearly fatal levels even through annual flares, reaching 1.32 Gy (8.09 Sv), and 1.09 Gy (6.68 Sv), respectively (see Figures \ref{fig:pf-4planet-Gy} and \ref{fig:pf-4planet-Sv}).

\subsection{Summary of XUV studies}
The following items are not well characterized in the presented models: (i)  the MUSCLES survey provides stellar spectra ranging from XUV to IR based using observed (Chandra \& XMM, and HST) and empirical estimates. However, the MUSCLES study includes stars earlier than M4 dwarfs ($T_{eff} >$ 3000 K), and thus there is no relevant data for cooler stars, such as TRAPPIST-1 whose $T_{\mathrm{eff}}$ is almost 2500 K. Our models based on these assumptions should be updated with new observations to be performed in the near future \citep{Airapetian2019}. (ii) The correlation between XUV fluxes and starspot sizes, assumed from previous solar observations, is currently only at a hypothetical stage especially for cool M-dwarfs. The specific relationship should be investigated in detail in accordance with photospheric temperature and wavelength. In short, we have to investigate each stellar temperature (spectral class) and planetary body in more detail to clarify those relationships. (iii) In general, there are no sufficient observational results about stellar activities especially for active M dwarfs, which should be the focus in future observations (iv) We need to deepen the survey for active M dwarfs in collecting more UV-EUV data, as most stellar objects determined by the MUSCLES survey are not active M dwarfs, except for Proxima Centauri. Moreover, as for Proxima Centauri \citep{Wargelin2017}, the stellar activity may change in accordance with the stellar cycle. (v) We made the first assumption of the ratio for XUV \& EUV in this survey, which should be supported and adopted through sufficient observational results. The Mega-MUSCLES project (\citet{Froning2018}) will also focus on a survey for TRAPPIST-1, which we can check the validity of our assumption. 

\section{Conclusion}
Our results suggest that both SPE and XUV fluxes doses are significantly higher 
at their TOA of close-in exoplanets around M dwarfs than those at Earth. For an exoplanet with a thick atmosphere (with an ozone layer for UV dose) the above extreme fluxes
do not affect the dose of ionizing radiation at the planetary surface. When a strong planetary magnetic field is not present or a stellar driver (stellar wind or a CME) is strong enough to perturb the global field and induce strong ionospheric currents that can dissipate into the heat (\citealt{Cohen2014}; \citealt{Airapetian2017b}), large XUV fluxes and massive winds can thus affect the erosion  of the atmosphere reducing its thickness on geological time scales (\citealt{Airapetian2017b}; \citealt{Garcia-Sage2017}). 
Accordingly, if the ozone layer is efficiently destroyed by SEP events and the exoplanetary atmosphere is eroded via atmospheric escape 
(\citealt{Segura2010}; \citealt{Airapetian2017a}; \citealt{Tilley2019}), 
then stellar XUV emission can penetrate into the planetary surface,
and provide detrimental conditions to complex lifeform. 
According to our scenario, if the atmospheric depth is \textless 1/10 of the terrestrial one, radiation doses become fatal for Proxima Centauri b and TRAPPIST-1 e even under Annual Maximum flare. 
Our new proposed scenario also suggests that under high energy and frequency of stellar flares, we can expect the impact of ionizing radiation on terrestrial-type life forms is not critical for their evolution if the atmosphere is thick enough ( $\approx$ 1 bar). Further efforts should explore the conditions for efficient removal of the UV protection layer (as Ozone layer) that prevents annual UV dose.
In this work, we developed a universal estimation method of the stellar flare frequency based on the starspot area observed on each host star. The stellar flare frequency varies from Annual Maximum Flare, 10-year flare, and possible maximum energy from the star. These estimations were based on \citet{Maehara2017} and \citet{Shibata2013}.

Our study indicates that for most ``habitable" planets orbiting M-class stars, the radiation dose at the TOA caused by periodic solar superflare activity is far in excess of the critical threshold for terrestrial life to survive. However, the simulations also indicate that planets with sufficient atmospheric thickness and density (N$_2$+O$_2$, CO$_2$ and H$_2$) are, at their crustal surfaces, protectively buffered from the radiation flux arriving at the TOA due to atmospheric attenuation of the incident radiation. This supports the hypotheses of Proxima Centauri b as a habitable planet once discovered; even though it is located very close to its host star.  In some cases, the incident dose may be reduced sufficiently to support surface life and this is most notable for atmospheric systems containing significant H$_2$ which would be more common for planets larger than the Earth.

In cases where atmospheric thickness alone is not sufficient to attenuate the radiation from solar superflares to non-life threatening levels, the enhanced attenuation efficacy presented by oceanic bodies may prove critical in preserving marine but not land-based lifeforms.
The critical atmospheric depth for each planet to secure terrestrial-type life form evolution on the surface of each planet are, 2.14 $\times 10^2$ g/cm$^2$ for the Proxima Cen b 4.68 $\times 10$ g/cm$^2$ for the Ross 128 b, and 2.04 $\times 10^2$ g/cm$^2$ for the TRAPPIST-1 e.
If we set the critical dose as 1 mSv per year, the critical atmospheric depth becomes, 1.05 $\times 10^3$ g/cm$^2$ for the Proxima Cen b, 7.8 $\times 10^2$ g/cm$^2$ for the Ross 128 b, and  1.04 $\times 10^3$g/cm$^2$ for the TRAPPIST-1 e. 

Do stellar flares present a common life-limiting factor for the development of complex lifeforms throughout the universe? Our simulations certainly suggest this might be the case, but equally, for close-in exoplanets around M-type stars that have no substantial atmosphere due to lack of degassing (from volcano-tectonic activity), the atmospheric depth might be insufficient to support terrestrial-type life in the first place, regardless of the periodic sterilizing effect of solar flares.

Future galactic surveys of extrasolar planets aimed at determining habitable planets, should characterize stellar activity as part of the analysis process and couple this with assessments of atmospheric composition, the presence of liquid water and the rate of planetary rotation versus superflare duration.  Indications of liquid water on any exoplanet surface, especially oceanic bodies, will likely be a critical indicator for habitability due to the effective protection it provides from solar radiation.

\acknowledgments

Authors express their sincere thanks to Astronomical Observatory, Kyoto University and NASA/GSFC by providing us relevant useful datasets and advice. 
Authors also express their sincere thanks to Prof. Takao Doi, Unit of the Synergetic Studies for Space, Kyoto University for his supports 
for ExoKyoto and observation projects, 
together with his sincere encouragements for our survey. 
Special thanks to Ms. Mayumi Tatsuda, secretary of Earth \& Planetary Water Resources Assessment Laboratory, 
Graduate School of Advanced Integrated Studies in Human Survivability for her support during compilation of the whole survey. 
This work was also supported by JSPS KAKENHI Grant Numbers JP16J00320, JP16J06887, JP16H03955, JP17H02865, JP17K05400, JP18J20048, JP18H01569 and MEXT grant number 26106006.
Vladimir Airapetian was supported by NASA grant 80NSSC17K0463, TESS Cycle 1 grant 80NSSC19K0381 and NASA/GSFC ISFM SEEC grant.

\appendix
\section{Method}
\subsection{Radiation dose}
As aforementioned, two types of radiation doses, the absorbed dose in Gy and the effective dose in Sv, were deduced from the simulation. In general, the absorbed doses are higher than the effective doses at the TOA because lower energy protons, which have a small impact on the human body due to their shorter range, predominantly contribute to the dose at the location. In contrast, the relation is reversed at the ground level because of the contribution of neutrons, which have a more significant impact on the human body, and become very important in such deeper locations.

In this study, we need to presume the critical dose for discussing the habitability on a planet. In general, mortality of radiation exposure is discussed in the absorbed dose throughout the whole body,
e.g., the whole-body absorbed dose that is lethal for half of the exposed individuals, LD$_{50}$ is around 4 Gy for photon exposure \citep{ICRP2007}.
However, LD$_{50}$  is expected to be different for cosmic-ray exposure due to higher relative biological effectiveness. In addition, the whole-body absorbed dose depends
on the size of the creature; the radiological sensitivities significantly vary with species. Thus, we decided to select the effective dose as an index for discussing the habitability because it is the most well known radiological protection quantity, and set the critical dose to 10 Sv per annually occurring Stellar Proton Events.

\subsection{Parameters and Equations}
A single flare event occurs on the stellar surface and the energy via electromagnetic wave radiates in all directions, whereas during a proton event, the energy has clear directionality. 
For the Solar Proton Events (SPEs), the release angle of the proton may be limited within a certain angle from the solar equator. We consider the total area, which may be affected by the SPEs that can be expressed as:
\begin{equation}
        A_{\mathrm{SPE}} = 2 \pi R_{\mathrm{e}}  \times T_{\mathrm{D}} \times H_{\mathrm{SPE}},
\label{eq:a_spe}
\end{equation}
\begin{equation}
        T_{\mathrm{D}} = 2 R_{\mathrm{e}} sin \theta_{\mathrm{V}},
\label{eq:t_d}
\end{equation}
\begin{equation}
        H_{\mathrm{SPE}} = \frac{\theta_{\mathrm{H}}}{360},
\label{eq:h_spe}
\end{equation}
in which $A_{\mathrm{SPE}}$: Total flare affected area in 1au distance, $R_{\mathrm{e}}$: Earth's Semimajor axis, $H_{\mathrm{SPE}}$: ratio of SPE horizontal release angle over entire orbital circle,  $\theta_{V}$ the vertical release angle of SPEs, $\theta_{H}$ the horizontal release angle of SPE, (in this study we assume that $\theta_{V}$ and $\theta_{H}$ are equal to 15 and 90 degrees, respectively).

Thus, the expected SPE energy per unit area received at Earth's TOA during a determined period (1 year) can be expressed as:
\begin{equation}
        E_{\mathrm{SPE_{Earth}} } = \frac{E_{\mathrm{flare}}\times H_{\mathrm{P}} \times R_{\mathrm{SPE}}}{ A_{\mathrm{SPE}}},
\label{eq:e_spe_earth}
\end{equation}
in which $H_{\mathrm{P}}=\theta_{\mathrm{H}}/180$: horizontal exposure probability (we employ 0.5), and $R_{\mathrm{SPE}}$: fraction of SPE energy in total flare energy per year. In this study, we employed 0.25 \citep{Aschwanden2017}.

From the above equation, we may calculate fluence on any exoplanet using the relative proportion to the fluence at Earth's TOA. 

\subsection{Normalization}
In order to apply this comparison with stellar flares, we employed a normalized value based on GOES X-ray class of the flare events. We can safely assume that total energy of flares can be estimated from the GOES X-ray class \citep{Namekata2017}. Accordingly, we assumed that the total energy of the Carrington-class event (estimated as X45 class) as $4.5 \times 10^{25}$ Joules , and GLE43 (estimated as X13 class) as $1.3 \times 10^{25}$ Joules . 

To integrate the above reference of Solar Proton Events, applicable for Stellar Proton Events, the following constants are determined as:
\begin{equation}
        E_{\mathrm{PE_{GLE43}} } = \frac{E_{\mathrm{GLE43}} \times R_{\mathrm{SPE}}}{ A_{\mathrm{SPE}}},
\label{eq:e_pe_gle43}
\end{equation}
\begin{equation}
        E_{\mathrm{PE_{Car}} } = \frac{E_{\mathrm{Car}}\times R_{\mathrm{SPE}}}{ A_{\mathrm{SPE}}},
\label{eq:e_pe_car}
\end{equation}
in which $E_{\mathrm{GLE43}}$: total energy of GLE43 (estimated as $1.3 \times 10^{32} (\mathrm{erg}) = 1.3 \times 10^{25} (\mathrm{Joule}) $ ), $E_{\mathrm{Car}}$: total energy of Carrington-class event (estimated as $4.5 \times 10^{32} (\mathrm{erg}) = 1.3 \times 10^{25} (\mathrm{Joule})$) . Note that coefficient $ H_{\mathrm{P}} $ has not been multiplied as it is obvious that those events released SEP toward the Earth.

\subsection{XUV Radiation}

XUV radiation is considered one type of harmful radiation released from a single stellar flare event. The proportion of XUV radiation in a single flare event has not been fully determined. However, according to \citet{Aschwanden2017}, the energetic portion of the UV continuum (with the sum of those ranges 200-228 $\mathrm{\AA}$, 370-504 $\mathrm{\AA}$, 504-912 $\mathrm{\AA}$, 1464-1609 $\mathrm{\AA}$, and 1600-1740 $\mathrm{\AA}$) is $3.96 \times 10^{29}$ erg, equivalent to $1.8 \%$ of estimated total flare energy (X2.2 class event $2.2 \times 10^{31}$ erg) in a solar flare. In this study, since there are no related studies about the proportion of UV radiation as a function of stellar temperature, diameter, and other factors; we use solar flare results \citep{Aschwanden2017} as a reference, and roughly assume that the UV portion of 10\% of total flare energy as the most extreme case. To determine the portion of XUV in the total UV dose of the flare as one extreme case we assume that 50\% of the total UV energy emitted during stellar flare is XUV energy.

When determining normal UV irradiation for each of the exoplanets, since there are no direct observations for most of the target stars, we synthesize the spectra using the published spectra curve from the MUSCLES project (\citealt{France2016}; \citealt{Youngblood2016}; \citealt{Loyd2016}) 
(using dapt-const-res-sed SED in version 2.2 dataset) and applied a similar spectral on the basis of temperature, trying to create an extreme UV case of a similar type to the star. We used the MUSCLES-observed spectra for 13 stars (GJ1214, GJ551(Proxima Cen), GJ876, GJ436, GJ581, GJ667C, GJ176, GJ832, HD85512, HD40307, HD97658, eps Eridanis, and the Sun) and synthesized the spectra for the target stars (TRAPPIST-1, Ross-128, Kepler-283, Kepler-1634). In the synthesizing process we divide the whole spectra into two, (i) one that is mostly related to the photospheric temperature, applied for IR, VR, and UVA \& UVB, in particular for wavelengths longer than 1200 $\mathrm{\AA}$; and (ii) one that is mostly related to the magnetic activity mainly in the chromosphere, applied for XUV, in particular for the wavelength from 1-1200 $\mathrm{\AA}$. For (ii) we made the hypotheses that the intensity of XUV is in proportion to the starspot area, based on the hypothesis that the starspot area represents the average magnetic field strength of the target star, thus chromospheric activity level. In this study, since our aim is to evaluate extreme cases under different star systems, we introduce a ratio to specify the weight of each effect. Currently, we set two extreme cases when synthesizing spectra of XUV portion, as (i) 100 \% of total-magnetic-flux-related (magnetic) term, generated from observed XUV (GJ551) calculated using relative portion of the starspot area and (ii) 60 \% of the magnetic term + 40 \% of the photosphere-temperature-related(photospheric) term. We set 20 \% of the magnetic term + 80 \% of the photospheric term when synthesizing the spectra of the EUV portion, while applying the following equation to synthesized fluxes in each wavelength: 
\begin{equation}
        F_{\mathrm{I} }= R_{\mathrm{A_{spotUV}}} \times F_{\mathrm{A_{spot}}} + (1 - R_{\mathrm{A_{spotUV}}} ) F_{\mathrm{T_{eff}}},
\label{eq:f1}
\end{equation}
in which  $F_{\mathrm{I} }$: UV flux for target stars,  $R_{\mathrm{A_{spotUV}}}$ : Impact ratio of starspot in UV(same for XUV),   $F_{\mathrm{A_{spot}}}$ :  flux calculated from the relative size of the starspot area, normalized by observed GJ551 (Proxima Cen) starspot area and its UV(XUV) spectra, and $F_{\mathrm{T_{eff}}}$ :  flux calculated from the photospheric temperature of the star. 

For example, when synthesizing a portion of IR, VR, FUV, MUV, \& NUV for TRAPPIST-1 (2550K), for the VR, IR region, we employed a spectra of GJ1214 (2935K) as a proxy for a lower temperature star. The spectra of Ross-128 (3192K)  was computed with GJ876(3062K) and GJ436(3281K). The spectra of Kepler-283(4351K) was calculated using HD85512(4305K) and HD40307(4783K), and the spectra of Kepler-1634(5474K) was synthesized with that of eps Eridanis and our Sun.  For the chromospheric component (XUV) we synthesized spectra mainly from Proxima Centauri and eps Eridanis, whose XUV components were much higher than that of other stars.

Then, we calculated the portion of UV (including XUV and other) and set its value as the radiation boundary from the central star. Irradiance at the top of the atmosphere of each planet can be calculated using normal radiation propagation. Estimating UV energy from stellar flares, we applied the statistical occurrence probability and possible total energy of the flare from each central star. Applying this UV energy ratio and projected planetary position, we calculated the energy at the TOA of each planet. Using the above hypotheses, the most intensive UV and XUV at the TOA induced by annual maximum flares, assuming this portion, were observed on Ross-128 b, reaching 8.75 $\times 10^{5} J/m^2$ for total UV and 4.38 $\times 10^{5} J/m^2$ for XUV (1-1200 $\mathrm{\AA}$ ), followed by Kepler-1634 b with 1.24 $\times 10^{5} J/m^2$, Proxima Cen b with 4.25 $\times 10^{4} J/m^2$, and TRAPPIST-1 d with 3.85 $\times 10^{4} J/m^2$, as shown in Table \ref{tab:Table2}. Although all values are largely compared with terrestrial impact values, they are all below 0.001\% of the terrestrial annual UV dose at the TOA (estimated as 4.31 $\times 10^{9} J/m^2$, shown in Table \ref{tab:Table2}. Also, by comparing the total dose during annual irradiation, the total value is far below the annual dose for the terrestrial case. The total annual UV (10-4000 $\mathrm{\AA}$)  dose does not reach the terrestrial level for all of the exoplanets, with the maximum value 0.56 in the terrestrial case of Kepler-1634 b. 

At the same time, limiting the dose of XUV(1-1200 $\mathrm{\AA}$) , most habitable exoplanets have a higher value compared with that of the Earth, reaching up to 76 times the value at Proxima Centauri b, followed by TRAPPIST-1 e with 65 times in the (ii) calculation, since the XUV portion among normal irradiation of the Sun is smaller than the observed M dwarf and thus it reflects an annual dose. On TRAPPIST-1 b, c and d, the XUV portion reaches a significant value of 418, 223, and 112 respectively (as shown in Table \ref{tab:Table4}). When applying the most extreme cases (i), TRAPPIST-1 e becomes the highest, reaching 100 times the value compared with Earth. On TRAPPIST-1 b, c and d for the (i) calculation, the XUV portion reaches significant values of 644, 344, and 173 times respectively (as shown in Table \ref{tab:Table3}).They are, however, generally considered as non-habitable planets in the system. Having relevant atmospheric depth and an ion layer, equivalent to that of the Earth- where most of XUV is absorbed by the factor of $10^{-15} - 10^{-22}$, the irradiation may be absorbed before reaching the planetary surface.  The flare impact on the UV dose only becomes important when superflares induce a significant reduction of the ozone layer \citep{Segura2010}, or when they accelerate significant atmospheric escape \citep{Airapetian2016}. Accordingly the atmospheric protection becomes more important for those planets due to higher ratio of normal irradiation of XUV.

\begin{table}[ht!]
\footnotesize
\centering
\renewcommand{\thetable}{\arabic{table}}
\caption{UV energy at TOA, with synthesized spectra assuming 100 percent from Starspot Area}
\label{tab:Table5}
\begin{tabular}{|l|c|c|c|c|c|c|c|c|c|c|c|c|c|c|}
\tablewidth{0pt}
\hline 
Exoplanet 
& $E^{\rm{flare}}_{\rm{UV}}$ 
& $\frac{E^{\rm{flare}}_{\rm{UV}}}{E^{\rm{flare}}_{\rm{UV, Earth}}}$ 
& $\frac{E^{\rm{flare}}_{\rm{UV}}}{E^{\rm{flux}}_{\rm{UV, Earth}}}$ 
& $E^{\rm{flare}}_{\rm{XUV}}$
& $\frac{E^{\rm{flare}}_{\rm{XUV}}}{E^{\rm{flare}}_{\rm{XUV, Earth}}}$ 
& $\frac{E^{\rm{flare}}_{\rm{XUV}}}{E^{\rm{flux}}_{\rm{XUV, Earth}}}$ 
& $E^{\rm{normal}}_{\rm{XUV}}$ 
\\
Name & [J m$^{-2}$] & &  [\%] & [J m$^{-2}$]& & [\%] & [J m$^{-2}$] \\
 & [1] & [2] & [3] & [4] 
 & [5] & [6] & [7] \\
\hline
Kepler-283 c & 8.24E+03 & 6.85E+04 & 2.44E$-$06 & 4.12E+03 & 6.85E+04 & 2.36E$-$02 & 1.01E+05  \\ 
Kepler-1634 b & 4.60E+04 & 3.83E+05 & 1.37E$-$05 & 2.30E+04 & 3.83E+05 & 1.32E$-$01 & 7.07E+05  \\ 
Proxima Cen b & 1.57E+04 & 1.31E+05 & 4.66E$-$06 & 7.86E+04 & 1.31E+05 & 4.51E$-$02 & 1.33E+07 \\ 
Ross-128 b & 8.75E+04 & 7.28E+05 & 2.60E$-$05 & 4.38E+04 & 7.28E+05 & 2.51E$-$01 & 1.38E+06  \\ 
TRAPPIST-1 b & 5.30E+04 & 4.41E+05 & 1.57E$-$05 & 2.65E+04 & 4.41E+05 & 1.52E$-$01 & 1.12E+08 \\ 
TRAPPIST-1 c & 2.83E+04 & 2.35E+05 & 8.39E$-$06 & 1.41E+04 & 2.35E+05 & 8.11E$-$02 & 5.99E+07  \\ 
TRAPPIST-1 d & 1.42E+04 & 1.18E+05 & 4.22E$-$06 & 7.12E+03 & 1.18E+05 & 4.08E$-$02 & 3.01E+07  \\ 
TRAPPIST-1 e & 8.24E+03 & 6.86E+04 & 2.45E$-$06 & 4.12E+03 & 6.86E+04 & 2.37E$-$02 & 1.75E+07  \\ 
TRAPPIST-1 f & 4.75E+03 & 3.96E+04 & 1.41E$-$06 & 2.38E+03 & 3.96E+04 & 1.36E$-$02 & 1.01E+07   \\ 
TRAPPIST-1 g & 3.22E+03 & 2.68E+04 & 9.54E$-$07 & 1.61E+03 & 2.68E+04 & 9.23E$-$03 & 6.81E+06   \\ 
TRAPPIST-1 h & 1.65E+03 & 1.37E+04 & 4.89E$-$07 & 8.24E+02 & 1.37E+04 & 4.73E$-$03 & 3.49E+06  \\ 
Sol d (Earth) & 3.70E$-$01 & 3.08E+00 & 1.10E$-$10 & 1.85E$-$01 & 3.08E+00 & 1.06E$-$06 & 1.74E+05 \\ 
Sol e (Mars) & 1.60E$-$01 & 1.33E+00 & 4.72E$-$11 & 7.96E$-$02 & 1.33E+00 & 4.57E$-$07 & 7.51E+04   \\ 
\hline
\hline
Exoplanet 
& $E^{\rm{normal}}_{\rm{UV}} $ 
& $E^{\rm{normal}}_{\rm{Visible}}$ 
& $E^{\rm{normal}}_{\rm{IR}}$ 
& $E^{\rm{flare+quiescent}}_{\rm{XUV}}$ 
& $\frac{E^{\rm{flare+quiescent}}_{\rm{XUV}}}{E^{\rm{flare+quiescent}}_{\rm{XUV, Earth}}}$ 
& $E^{\rm{flare+quiescent}}_{\rm{UV}}$ 
& $\frac{E^{\rm{flare+quiescent}}_{\rm{UV}}}{E^{\rm{flare+quiescent}}_{\rm{UV, Earth}}}$ 
\\
Name & [J m$^{-2}$]& [J m$^{-2}$] & [J m$^{-2}$] & [J m$^{-2}$] & & [J m$^{-2}$] &  
\\
& [8]  & [9] & [10] & [11] & [12] & [13] & [14] \\
\hline
Kepler-283 c & 1.96E+09 & 1.12E+10 & 1.19E+10 & 1.05E+05 & 0.60 & 1.96E+09 & 0.58 \\ 
Kepler-1634 b & 1.90E+09 & 1.21E+10 & 1.41E+10 & 7.30E+03 & 4.19 & 1.90E+0.9 & 0.56 \\ 
Proxima Cen b & 2.44E+07 & 8.78E+08 & 2.74E+10 & 1.33E+07 & 76.21 & 2.44E+07 & 0.01 \\ 
Ross-128 b & 9.24E+07 & 5.01E+09 & 5.81E+10 & 1.42E+06 & 8.17 & 9.25E+07 & 0.03 \\ 
TRAPPIST-1 b & 2.53E+08 & 8.97E+09 & 1.72E+11 & 1.12E+08 & 644.10 & 2.53E+08 & 0.07\\ 
TRAPPIST-1 c & 1.35E+08 & 4.79E+09 & 9.19E+10 & 5.99E+07 & 343.66 & 1.35E+08 & 0.04   \\ 
TRAPPIST-1 d &  6.79E+07 & 2.41E+09 & 4.63E+10 & 3.01E+07 & 172.95 & 6.79E+07 & 0.02  \\ 
TRAPPIST-1 e &  3.93E+07 & 1.40E+09 & 2.68E+10 & 1.75E+07 & 100.19 & 3.97E+07 & 0.01 \\ 
TRAPPIST-1 f &  2.27E+07 & 8.04E+08 & 1.54E+10 & 1.01E+07 & 57.76 & 2.27E+07 & 0.01  \\ 
TRAPPIST-1 g &  1.53E+07 & 5.44E+08 & 1.05E+10 & 6.81E+06 & 39.09 & 1.53E+07 & 0.00  \\ 
TRAPPIST-1 h & 7.86E+06 & 2.79E+08 & 5.36E+09 & 3.49E+06 & 20.03 & 7.86E+06 & 0.00  \\ 
Sol d (Earth) &  3.37E+09 & 1.93E+10 & 2.04E+10 & 1.74E+05 & 1.00 & 3.37E+09 & 1.00 \\ 
Sol e (Mars) &  1.45E+09 & 8.32E+09 & 8.81E+09 & 750E+04 & 0.43 & 1.45E+09 & 0.43  \\ 
\hline
\multicolumn{4}{l}{[1] UV Energy by Annual Maximum Flare at TOA }&
\multicolumn{4}{l}{[2] Ratio to Earth's Annual Maximum Flare}\\
\multicolumn{4}{l}{[3] Ratio to Earth's annual UV flux at TOA }&
\multicolumn{4}{l}{[4] XUV Energy by Annual Maximum Flare at TOA} \\
\multicolumn{4}{l}{[5] Ratio to Earth's Annual Maximum Flare }&
\multicolumn{4}{l}{[6] Ratio to Earth's annual UV flux at TOA}\\
\multicolumn{8}{l}{[7] Annual XUV Energy by Normal Stellar Radiation at TOA }\\
\multicolumn{8}{l}{[8] Annual UV Energy by Normal Stellar Radiation at TOA}\\
\multicolumn{8}{l}{[9] Annual Visible Ray Energy by Normal Stellar Radiation at TOA} \\
\multicolumn{8}{l}{[10] Annual IR Energy by Normal Stellar Radiation at TOA }\\
\multicolumn{8}{l}{[11] Annual Total (flare + Quiescent) XUV Energy at TOA}\\
\multicolumn{8}{l}{[12] Ratio to Earth / Annual Total (flare + Quiescent) XUV Energy at TOA}  \\
\multicolumn{8}{l}{[13] Annual Total  (flare + Quiescent) UV Energy at TOA} \\
\multicolumn{8}{l}{[14] Ratio to Earth of Annual Total  (flare + Quiescent)UV Energy at TOA}\\
\multicolumn{8}{l}{TOA - Top of Atmosphere ($\approx$ 0 g/cm$^2$)}
\end{tabular}
\end{table}


\begin{thebibliography}{}

\bibitem[Airapetian et al.(2016)]{Airapetian2016} Airapetian, V.~S., Glocer, A., Gronoff, G., H{\'e}brard, E., \& Danchi, W.\ 2016, Nature Geoscience, 9, 452 
\bibitem[Airapetian et al.(2017a)]{Airapetian2017a} Airapetian, V. S., Glocer, A., Khazanov, G. V., et al.\ 2017a, \apjl, 836, L3 
\bibitem[Airapetian et al.(2017b)]{Airapetian2017b} Airapetian, V. S., Jackman, C., Mlynzcak, M., Danchi, W.,\& Hunt, M. ,\ 2017b, Scientific Reports, 7, 14141
\bibitem[Airapetian et al.(2019)]{Airapetian2019} Airapetian, V. et al.\ 2019, International Journal of Astrobiology, eprint arXiv:1905.05093
\bibitem[Aschwanden et al.(2017)]{Aschwanden2017} Aschwanden, M.~J., Caspi, A., Cohen, C.~M.~S., et al.\ 2017, \apj, 836, 17 
\bibitem[Atri(2017)]{Atri2017} Atri, D.\ 2017, \mnras, 465, L34 
\bibitem[Cohen et al.(2014)]{Cohen2014} Cohen, O., Drake, J.~J., Glocer, A., et al.\ 2014, \apj, 790, 57
\bibitem[Davenport(2016)]{Davenport2016} Davenport, J.~R.~A.\ 2016, \apj, 829, 23
\bibitem[Elkins-Tanton and Seafer (2008)]{Elkins2008} Elkins-Tanton, Seager, S., \ 2008 \apj 685, 1237-1246
\bibitem[France et al.(2016)]{France2016} France, K., Loyd, R.~O.~P., Youngblood, A., et al.\ 2016, \apj, 820, 89 
\bibitem[Froning et al.(2018)]{Froning2018} Froning, C.~S., France, K., Loyd, R.~O.~P., et al.\ 2018, American Astronomical Society Meeting Abstracts \#231, 231, 111.05 
\bibitem[Garcia-Sage et al.(2017)]{Garcia-Sage2017} Garcia-Sage, K., Glocer, A., Drake, J.~J., Gronoff, G., \& Cohen, O.\ 2017, \apjl, 844, L13 
\bibitem[Gopalswamy et al.(2017)]{Gopalswamy2017} Gopalswamy, N., Yashiro, S., Akiyama, S., \& Xie, H.\ 2017, \solphys, 292, 65 
\bibitem[Grie{\ss}meier et al.(2015)]{Grie2015} Grie{\ss}meier, J.-M., Tabataba-Vakili, F., Stadelmann, A., Grenfell, J.~L., \& Atri, D.\ 2015, \aap, 581, A44 
\bibitem[ICRP(2007)]{ICRP2007} Intenational Commission on Radiological Protection (ICRP),\ 2007, ICRP Publication, 103, Ann. ICRP 37 (2-4)
\bibitem[ICRP(2010)]{ICRP2010} International Commission on Radiological Protection (ICRP), \ 2010, ICRP Publication, 116, Ann. ICRP 40 (2-5)
\bibitem[Jakosky et al.(2015)]{Jakosky2015} Jakosky, B.~M., Grebowsky, J.~M., Luhmann, J.~G., et al.\ 2015, Science, 350, 0210 
\bibitem[Kasting(1988)]{Kasting1988} Kasting, J.~F.\ 1988, \icarus, 74, 472 
\bibitem[Kasting et al.(1993)]{Kasting1993} Kasting, J.~F., Whitmire, D.~P., \& Reynolds, R.~T.\ 1993, \icarus, 101, 108 
\bibitem[Kay et al.(2016)]{Kay2016} Kay, C., Opher, M., \& Kornbleuth, M.\ 2016, \apj, 826, 195 
\bibitem[Kopparapu et al.(2013)]{Kopparapu2013} Kopparapu, R.~K., Ramirez, R., Kasting, J.~F., et al.\ 2013, \apj, 765, 131 
\bibitem[Kumari et al.(2017)]{Kumari2017} Kumari, A., Ramesh, R., Kathiravan, C., \& Gopalswamy, N.\ 2017, \apj, 843, 10 
\bibitem[Lammer et al.(2018)]{Lammer2018} Lammer, H., Zerkle, A.,L., Gebauer, S. et al. \ 2018{\"o} A\&ArV, 26, 72

\bibitem[Lingam \& Loeb(2017)]{Lingam2017} Lingam, M., \& Loeb, A.\ 2017, \apj, 848, 41 
\bibitem[Loyd et al.(2016)]{Loyd2016} Loyd, R.~O.~P., France, K., Youngblood, A., et al.\ 2016, \apj, 824, 102 
\bibitem[Maehara et al.(2012)]{Maehara2012} Maehara, H., Shibayama, T., Notsu, S., et al.\ 2012, \nat, 485, 478 
\bibitem[Maehara et al.(2015)]{Maehara2015} Maehara, H., Shibayama, T., Notsu, Y., et al.\ 2015, Earth, Planets, and Space, 67, 59 
\bibitem[Maehara et al.(2017)]{Maehara2017} Maehara, H., Notsu, Y., Notsu, S., et al.\ 2017, \pasj, 69, 41
\bibitem[Mclean et al.(2017)]{Mclean2017} McLean, A.,R., et al.\ 2017,  Proc. R. Soc., B, 284, 20171070 
\bibitem[Miyake et al.(2012)]{Miyake2012} Miyake, F., Nagaya, K., Masuda, K., \& Nakamura, T.\ 2012, \nat, 486, 240 
\bibitem[Miyake et al.(2013)]{Miyake2013} Miyake, F., Masuda, K., \& Nakamura, T.\ 2013, Nature Communications, 4, 1748 
\bibitem[Namekata et al.(2017)]{Namekata2017} Namekata, K., Sakaue, T., Watanabe, K., et al.\ 2017, \apj, 851, 91 
\bibitem[Nitta et al.(2012)]{Nitta2012} Nitta, N.~V., Liu, Y., DeRosa, M.~L., \& Nightingale, R.~W.\ 2012, \ssr, 171, 61 
\bibitem[Notsu et al.(2013)]{Notsu2013} Notsu, Y., Shibayama, T., Maehara, H., et al.\ 2013, \apj, 771, 127 
\bibitem[Notsu et al.(2015a)]{Notsu2015a} Notsu, Y., Honda, S., Maehara, H., et al.\ 2015a, \pasj, 67, 32 
\bibitem[Notsu et al.(2015b)]{Notsu2015b} Notsu, Y., Honda, S., Maehara, H., et al.\ 2015b, \pasj, 67, 33
\bibitem[Notsu et al.(2019)]{Notsu2019} Notsu, Y., Maehara, H., Honda, S., et al.\ 2019, \apj, in press (arXiv:1904.00142)
\bibitem[Ramirez et al.(2019)]{Ramirez2019} Ramirez, R., M., et al. \ 2019 Astro2020, in press (arXiv:1903.03706)
\bibitem[Ribas et al.(2017)]{Ribas2017} Ribas, I., Gregg, M.~D., Boyajian, T.~S., \& Bolmont, E.\ 2017, \aap, 603, A58 
\bibitem[Sato et al.(2014)]{Sato2014} Sato, T., Kataoka, R., Yasuda, H., et al. \ 2014, Radiation Protection Dosimetry, 161, 274
 \bibitem[Sato et al.(2015)]{Sato2015} Sato, T. ,\ 2015, PLoS ONE 10(12): e0144679  
\bibitem[Sato et al.(2018a)]{Sato2018a}
Sato, T.,  Iwamoto, Y., Hashimoto, S., et al.,\ 2018,  Journal of Nuclear Science and Technology, 55, 6, 684 
\bibitem[Sato et al.(2018b)]{Sato2018b}
Sato, T., Kataoka, R., Shiota, D., et al.,\ 2018, Space Weather 
\bibitem[Schrijver et al.(2015)]{Schrijver2015} Schrijver, C.,J.et al. \ 2015, Advanced in Space Research 55 (2745-2807)
\bibitem[Segura et al.(2010)]{Segura2010} Segura, A., Walkowicz, L.~M., Meadows, V., Kasting, J., \& Hawley, S.\ 2010, Astrobiology, 10, 751 
\bibitem[Shibayama et al.(2013)]{Shibayama2013} Shibayama, T., Maehara, H., Notsu, S., et al.\ 2013, \apjs, 209, 5
\bibitem[Shibata et al.(2013)]{Shibata2013} Shibata, K., Isobe, H., Hillier, A., et al.\ 2013, \pasj, 65, 49 
\bibitem[Smart et al.(2006)]{Smart2006} Smart, D.~F., Shea, M.~A., \& McCracken, K.~G.\ 2006, Advances in Space Research, 38, 215 
\bibitem[Takahashi et al.(2016)]{Takahashi2016} Takahashi, T., Mizuno, Y., \& Shibata, K.\ 2016, \apjl, 833, L8 
\bibitem[Tilley et al.(2019)]{Tilley2019} Tilley, M., A., Segura, A., Meadows, V., Hawley, S.,\& Davenport, J.,\ 2019, Astrobiology, 19, 1
\bibitem[Townsend et al.(2006)]{Townsend2006} Townsend, L.~W., Stephens, D.~L., Hoff, J.~L., et al.\ 2006, Advances in Space Research, 38, 226 
\bibitem[UNSCEAR(2000)]{UNSCEAR2000}UNSCEAR 2000 Report, Vol. I 
\bibitem[Usoskin et al.(2013)]{Usoskin2013} Usoskin, I.~G., Kromer, B., Ludlow, F., et al.\ 2013, \aap, 552, L3 
\bibitem[Wargelin et al.(2017)]{Wargelin2017} Wargelin, B.~J., Saar, S.~H., Pojma{\'n}ski, G., Drake, J.~J., \& Kashyap, V.~L.\ 2017, \mnras, 464, 3281 
\bibitem[Xapsos et al. (2000)]{Xapsos2000} 
Xapsos, M.A., Barth, J., L., Stassinpoulos, E., et al. ,\ 2000, IEEE Transactions on Nuclear Science, 47, 6
\bibitem[Youngblood et al.(2016)]{Youngblood2016} Youngblood, A., France, K., Loyd, R. O. P.,  et al.,\ 2016, \apj, 824, 101
\end{thebibliography}
\end{document}